\documentclass[journal=pasa, manuscript=research-paper, year=2025, volume=?]{cup-journal}
\usepackage{hyperref}
\hypersetup{
    colorlinks = true,
    allcolors  = {blue} 
}

\newif\ifshowrevisions
\showrevisionstrue   

\usepackage{amssymb}
\usepackage{amsmath}
\usepackage[nopatch]{microtype}
\usepackage{booktabs}
\usepackage{natbib}
\definecolor{darkgreen}{rgb}{0.0, 0.5, 0.0}
\newcommand{\msun}[1]{M$_\odot$}

\title{Second Generation Planet Formation in Post-AGB Discs: Testing the Role of Gravitational Instability}

\author{Ali Pourmand}
\affiliation{School of Mathematical and Physical Sciences, Macquarie University, Sydney, NSW, Australia}
\alsoaffiliation{Astrophysics and Space Technologies Research Centre, Macquarie
University, Sydney, NSW, Australia}
\author{Devika Kamath} 
\affiliation{School of Mathematical and Physical Sciences, Macquarie University, Sydney, NSW, Australia}
\alsoaffiliation{Astrophysics and Space Technologies Research Centre, Macquarie
University, Sydney, NSW, Australia}
\author{Orsola De Marco}
\affiliation{School of Mathematical and Physical Sciences, Macquarie University, Sydney, NSW, Australia}
\alsoaffiliation{Astrophysics and Space Technologies Research Centre, Macquarie
University, Sydney, NSW, Australia}
\author{Mark Wardle}
\affiliation{School of Mathematical and Physical Sciences, Macquarie University, Sydney, NSW, Australia}
\alsoaffiliation{Astrophysics and Space Technologies Research Centre, Macquarie
University, Sydney, NSW, Australia}
\email[Ali Pourmand]{ali.pourmand@hdr.mq.edu.au}


\begin{document}

\begin{abstract}

Post-asymptotic giant branch (post-AGB) binary stars are evolved systems that host circumbinary discs formed through mass loss during late stage binary interactions. Their structural, morphological, kinematic, and chemical similarities to planet-forming discs suggest that these systems may act as sites of “second generation” planet formation. In this study, we assess whether the disc instability mechanism -- a proposed pathway for rapid, giant planet formation in some protoplanetary discs - can operate in post-AGB discs; motivated by their short lifetimes ($10^{4-5}$years). Using the Toomre criterion under well motivated assumptions for disc structure and size, mass, and thermal properties, we assess the conditions for gravitational instability. We first benchmark our analytical framework using well studied protoplanetary disc systems (including HL Tauri, Elias 2-27, GQ Lupi) before applying the same analysis to observed post-AGB discs. We find that post-AGB discs are generally gravitationally stable at present, due primarily to their low masses. Using viscous disk theory, we find that the discs were stable against collapse even in the past, when their masses were potentially higher. In contrast, several protoplanetary discs analysed in the same way show that they likely experienced gravitationally unstable phases early on. We also find that higher viscosity parameters ($\alpha \sim 10^{-2}$) are better aligned with expected post-AGB disc lifetimes. Finally, we revisit the planet formation scenario proposed for the post-common envelope system NN~Ser, first carried out by Schleicher and Dreizler and we show that gravitational instability could be feasible under specific, high disc mass assumptions. Overall, our results provide the first systematic theoretical assessment of gravitational instability in post-AGB discs, demonstrating that this mechanism is unlikely to dominate second generation planet formation in these systems, and underscoring the need to explore alternative pathways - such as core accretion - in future studies.
\end{abstract}

\section{Introduction}
\label{sec:introduction}
The {\it Kepler} and {\it Gaia} space telescopes have provided precise planet and stellar parameters for thousands of exoplanet systems \citep{2010Borucki,2018Thompson, 2016Gaia,2020berger} and high angular resolution observations with the Atacama Large Millimeter/ sub-millimeter Array (ALMA) have revealed direct signatures of ongoing planet formation in young stellar object (YSO) discs, such as gaps, rings, cavities, and spirals - as seen in systems like PDS\,70 \citep{Keppler_2019}, HL\,Tauri \citep{Booth_2020}, AB Aurigae \citep{Boccaletti_2020}, Elias 2-27 \citep{Perez_2016}, and HD 169142 \citep{Fedele_2017}.

Planets have also been detected or suspected around evolved stars. Notably, the first confirmed exoplanets were found orbiting the pulsar PSR B1257+12 - a system now known to host at least three planets \citep{1992Wolszczan}. The formation and/or survival pathway leading to planets existing in these extreme environments remains uncertain. For instance, \citet{1993Podsiadlowski} proposed that the disc from which pulsar planets formed originated through channels distinct from those forming YSO discs. Another example is the L2 Puppis system, an asymptotic giant branch (AGB) star with a circumbinary disc and a candidate planetary companion - offering a glimpse into a potential future for our own solar system \citep{Kervella_2016}. Similarly, NN~Ser, a post-common envelope binary, exhibits eclipse timing variations interpreted as evidence for two circumbinary planets \citep{Beuermann_2010} (though alternative explanations for these variations such as magnetic activity cycles or other dynamical effects remain under debate \citep{1992Applegate,Navarrete_2018,_zd_nmez_2023}). These planets have been proposed to form from gas falling back following a common envelope phase \citep{2014schleicher}, as their low masses suggest that they may not have survived direct interaction with the RGB/AGB progenitor. . Systems such as these point to the intriguing possibility of planet formation in previously unanticipated post-main sequence environments. 

Of specific interest are the discs around post-asymptotic giant branch (post-AGB) binaries. Post-AGB binaries are systems with a post-AGB star in the late stages of stellar evolution. The spectral energy distributions (SEDs) of the majority of post-AGB binary stars show a near-IR excess indicative of a circumbinary disc \citep{2006deRuyter,2014Kamath,Kamath_2015,Kluska_2022}. The formation mechanisms of these discs are still poorly understood, but they are presumed to have formed from late stage binary interactions, with more than 80 such discs now catalogued \citep[e.g.,][]{kamath_2019,Kluska_2022,Andrych_2023}. The observational properties of these discs have been extensively studied and reveal striking similarities to those of protoplanetary discs around YSOs. For instance, interferometric maps with ALMA and observations of $^{12}$CO and $^{13}$CO lines \citep{Bujarrabal_2015,Gallardo_Cava_2023} have shown that many of these discs exhibit stable (quasi)-Keplerian dynamical properties \citep{2006deRuyter,Gallardo_Cava_2021}. They also show inner cavities in the dust component of the disc that is comparable to the dust sublimation radius in many of these systems \citep{Kluska_2019,Corporaal_2023}. Furthermore, there is evidence of grain growth and dust grain evolution comparable to that seen in protoplanetary discs \citep{2005deruyter,Gielen_2011,Andrych_2023}. Moreover, detailed chemical abundance studies of post-AGB stars in binary systems reveal a photospheric chemical depletion of refractory elements \citep{Kluska_2018}, which has been interpreted as possibly being caused by a planet forming a dust-trapping gap in these discs \citep{Kluska_2022}. Photospheric depletion is stronger in post-AGB discs categorised as "transition discs" \citep{mohorian2025tracingchemicaldepletionevolved}, supporting the possibility that these discs harbour planets. Finally, the morphology of several of these transition post-AGB discs includes features such as cavities \citep{Corporaal2023}, with inner radii of up to 7.5 times the sublimation radius. These similarities have led to the hypothesis that post-AGB discs may also support planet formation.

On the other hand, key differences distinguish post-AGB discs from their YSO counterparts. Most notably, post-AGB discs have shorter estimated lifetimes of around $10^5$ years \citep[e.g.,][]{Bujarrabal_2017}, compared to $10^6$ years for typical YSO discs. They also tend to be lower in total mass (gas plus dust), typically ranging between $0.02$–$0.05$\,M$\odot$, whereas YSO discs span a broader range, with disc masses from $10^{-5}$ to 0.1\,M$\odot$ \citep{2023Bae,2023Manara}. Other differences include the presence of an evolved stellar radiation field, binary-driven disc dynamics, and differing initial conditions for the dust and gas content. 

Planets around main sequence stars are presumed to form primarily via core accretion or gravitational instability (\citealt{drazkowska2023planetformationtheoryera}). Core accretion is a bottom-up process involving the growth of small, grain-sized solids through coagulation and sticking during collisions and other mechanisms, eventually forming planetesimals and planetary embryos that can accrete additional material through gravitational interactions. The alternative theory, gravitational instability, is a top-down process in which massive regions of the disc undergo gravitational collapse, forming bound clumps that can contract and accrete material until planet-sized bodies emerge. This mechanism - similar in nature to the process of star formation - has been explored in seminal works such as \citet{1997boss,2002mayer}. Gravitational instability is typically invoked to explain planet formation in the outer regions of discs, where cooling timescales and disc mass conditions may favour instability and fragmentation. 

Gravitational instability can lead to planet formation on timescales comparable to the orbital period of the protoplanetary clump, making it a promising candidate for forming planets within short-lived post-AGB discs with lives of up to $10^5$ years \citep[e.g.][]{Bujarrabal_2016}. However, this mechanism also has limitations; most notably, it requires a sufficiently massive disc with the mass concentrated in a compact region for fragmentation to occur. Nevertheless, \cite{2014schleicher} applied this mechanism to the NN Ser system, arguing that a relatively massive and compact disc -- left behind after the CE phase -- could have undergone gravitational fragmentation. Their model produced planet masses comparable to those observed in NN Ser, supporting the plausibility of second generation planet formation via disc instability in this case.

In this study, we carry out a thorough analytical analysis to ascertain the possibility of planet formation in post-AGB discs through gravitational instabilities, while leaving core accretion to be investigated in a future study. We have anchored our analysis in a range of YSO systems for which planets, or planet-indicative disc features are reported. 

This paper is structured as follows. In Section \ref{sec:methods}, we introduce the analytical framework used to assess the possibility of planet formation via gravitational instability. In Section \ref{sec:ysos}, we apply this framework to a sample of YSO discs that are known or suspected to host planets or exhibit signatures of gravitational instability, as an initial test of the theory. Section \ref{sec:postagbs} presents the core focus of this work - an investigation of whether typical post-AGB discs can be gravitationally unstable. In Section \ref{sec:earlylife}, we explore how the masses and sizes of present-day discs may have evolved over time and assess their early-life stability. Section \ref{sec:nnser} revisits the analysis by \citet{2014schleicher} of the NN Ser system, where we attempt to reproduce their calculations using more accurate assumptions about the disc's temperature distributions. 

\section{Analytical Framework to Investigate Gravitational Instabilities}
\label{sec:methods}
In this section, we outline the theoretical framework we have used to evaluate the conditions under which a disc may become gravitationally unstable and fragment. We describe the Toomre stability criterion, cooling efficiency conditions, and various models for the disc surface density and thermal structure. These provide the basis for evaluating whether a given disc structure is susceptible to planet formation by fragmentation.

Gravitational instabilities occur in discs if the self-gravity of the disc is able to overcome the effects of thermal pressure and rotation. This is quantified by the Toomre parameter $Q$  \citet{1964Toomre}: 

\begin{equation}
\label{eq:toomre}
Q(r)=\frac{c_{\rm s}(r)\Omega(r)}{\pi G \Sigma(r)} \leq 1
\end{equation}
where $c_{\rm s}$ is the sound speed,  $\Omega$ 
is the Keplerian angular frequency of the disc, $\Sigma$ is the disc's surface density, $G$ is the gravitational constant 
, and $r$ is the radial distance from the center of the disc. While the classical Toomre criterion requires $Q \leq 1$ for instability against axisymmetric perturbations, simulations indicate that non-axisymmetric instabilities and spiral structures can emerge for $Q \leq 1.4$–1.7 \citep{Helled_2014}.
The sound speed is related to the temperature, $T(r)$, by the following expression:

\begin{equation}
\label{eq:sound_speed}
c_{\rm s}(r)=\sqrt{\eta \frac{k_{\rm B}T(r)}{\mu m_{\rm p}}},
\end{equation}
where $k_{\rm B}$ and $m_{\rm p}$ are the Boltzmann constant and proton mass, respectively, and we have assumed the adiabatic index $\eta=1$. \footnote{We denote the adiabatic index by $\eta$ to avoid confusion with the common use of $\gamma$ as a viscosity exponent in disc models.} We adopt a mean molecular weight of $\mu=2$ corresponding to molecular hydrogen gas.

For the optically thick regions of the disc: 
\begin{equation}
\label{eq:Tchiang}
T(r) = \left(\frac{\alpha}{4}\right)^{1/4}\left(\frac{R_{\rm star} }{r}\right)^{1/2}T_{\rm star} 
\end{equation}
\citep{Chiang_1997}, where $\alpha\approx 0.4R_{\rm star} /r$ is the grazing angle of the flux of stellar radiation incident on the disc, and is typically much smaller than unity, $R_{\rm star}$ is the photospheric radius of the star and $T_{\rm star}$ is the surface temperature of the star. This gives $T\propto r^{-0.75}$. This temperature profile is valid for optically thick discs in radiative equilibrium. 

Using this, we propose two temperature profiles for two regions inside and outside of the sublimation radius defined as:

\begin{equation}
    \label{eq:r_subl}
    R_{\rm subl} =\frac{1}{2}R_{\rm star} \left(\frac{T_{\rm star}}{T_{\rm subl}} \right)^2,
\end{equation}
where we assume $T_{\rm subl} =1200$~K is the dust sublimation temperature in post-AGB discs \citep{Kluska_2019}. Assuming radiative equilibrium, the temperature of the disc in the midplane will be
\begin{equation}
\label{eq:T_profile}
\begin{aligned}
T(r) = T_{\rm star} \left(\frac{R_{\rm star} }{2r}\right)^{0.5}
\, \text{for} \, r<R_{\rm subl} ,\\
T(r) =  \frac{A}{r^{0.75}}\, \text{for} \, r>R_{\rm subl} .
\end{aligned}
\end{equation}
where $A$ is a constant that can be calculated by making use of the fact that the two equations should predict the same temperature, $T_{\rm subl}$, at $r=R_{\rm subl}$:
\begin{equation}
    \label{eq:T_constant}
    A= T_{\rm subl}R_{\rm subl}^{0.75}.
\end{equation}
Using this we have: 
\begin{equation}
\label{eq:Tprofile}
\begin{aligned}
T(r) = T_{\rm star} \left(\frac{R_{\rm star} }{2r}\right)^{0.5}
\, \text{for} \, r<R_{\rm subl} ,\\
T(r) = T_{\rm subl} \left(\frac{R_{\rm subl}}{r}\right)^{0.75}\, \text{for} \, r>R_{\rm subl} .
\end{aligned}
\end{equation}
\subsection{Cooling Efficiency
}
\label{sec:cooling}
In order for gravitational instabilities to lead to disc fragmentation, the fragments must cool sufficiently rapidly such that thermal pressure does not halt the collapse. 
This criterion is expressed by the following inequality \citep{2001gammie}:
\begin{equation}
\label{eq:cooling_criterion}
    t_{\rm cool}<\frac{\beta}{\Omega},
\end{equation}
where $t_{\rm cool}$ is the cooling time and $\beta\approx3$ is a constant. It has been shown in simulations by \citet{2002mayer} that protoplanetary discs with minimum Toomre parameters of $Q_{\rm min}\sim 1.4$ start forming spiral structures which can fragment and form several planets within $\sim  1000$ years, if the cooling is efficient enough. The ratio of the thermal energy of the disc to its emissivity can be used to estimate cooling times  \citep{Rafikov_2005,Kratter_2010}:
\begin{equation}
\label{eq:coolingt}
    t_{\rm cool} = \frac{ c_s^2 \Sigma}{(\eta -1)}\frac{f(\tau)}{\sigma T^4},
\end{equation}
where $\sigma$ is the Stefan–Boltzmann constant, and $f(\tau)=\tau+1/\tau$ is a function to quantify the efficiency of disc cooling, which takes the vertical optical depth to be $\tau=\kappa \Sigma/2$, where $\kappa$ is the opacity.

\subsection{The Surface Density Profile}
\label{ssec:the_surface_density_profile}

\begin{figure}[hbt!]
\centering
 \includegraphics[width=\linewidth]{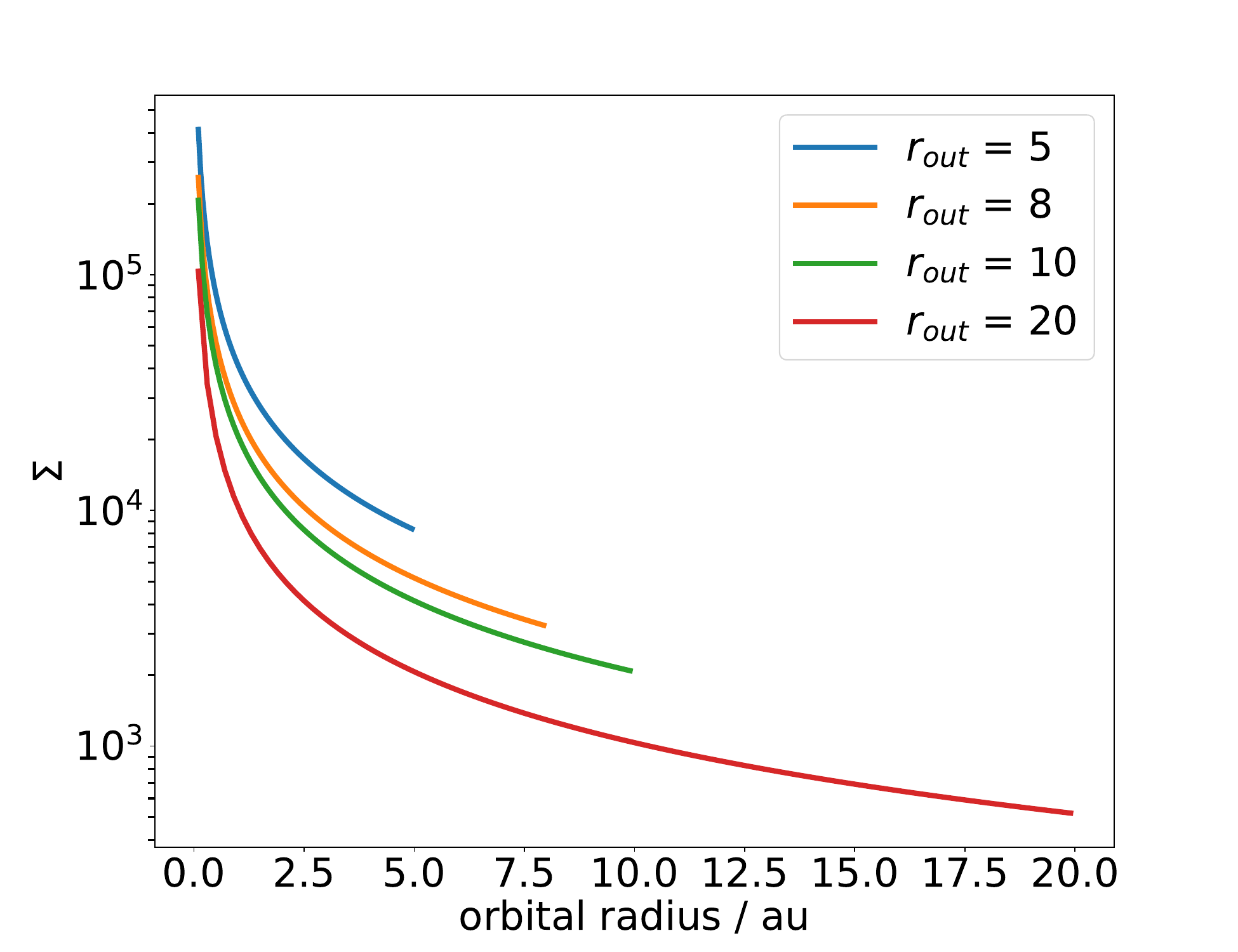}
\caption{Power-law surface density functions with $n=1$ as a function of orbital distance for $M_{\rm disc}=0.15~M_\odot$ for different values of $r_{\rm out}$, to demonstrate why the surface density is highly sensitive to $r_{\rm out}$.
}
\label{fig:sigmavsr_Q_1}
\end{figure}

The surface density profile, $\Sigma(r)$, plays a key role in determining the gravitational stability of the disc. A commonly adopted empirical form is a truncated power-law \citep{2014schleicher}:

\begin{equation}
    \Sigma(r) = \Sigma_0\left(\frac{r_{\rm out}}{r}\right)^{n},
\label{eq:surface_density_power_law}
\end{equation}
where  $r$ is the orbital distance, $r_{\rm out}$ is the outer disc's radius,  $n$ is the power law index that determines the steepness of the surface density profile and $\Sigma_0$ is a normalisation constant set by the disc's mass, $M_{\rm disc}$, $n$, and the outer radius:

\begin{equation}
    \label{eq:surfacedensity0schleicher}
\Sigma_0 =
\frac{(2-n)M_{\rm disc} }{2\pi r_{\rm out} ^{2}},
\end{equation}
which is only valid for $n<2$. This type of function is used to fit the surface densities of both YSO discs \citep{2023Speedie}, and post-AGB discs \citep{Corporaal_2023}. For a fixed disc mass, decreasing $r_{\rm out}$ increases the overall surface density, effectively making the disc more compact (Figure~\ref{fig:sigmavsr_Q_1}). 

In Section~\ref{sssec:surface_density_profiles_of_viscous_discs}, we introduce an alternative surface density formulation that addresses some of the limitations of this parametrisation, including the assumption of a sharply defined outer edge.

\subsubsection{Discs with Inner Cavities}

Post-AGB discs have inner cavities, evacuated regions, or holes centred on the central binary. These are not considered by Equation~\ref{eq:surface_density_power_law}. Moreover, Equation~\ref{eq:surfacedensity0schleicher} gives negative surface density values for power law indices $n\ge2$ . To account for this if our disc has a cavity, we use Equation~\ref{eq:surfacedensity0schleicher} but set $\Sigma(r)=0$ for $r < r_{\rm in} $, where $r_{\rm in} $ is the inner radius of the disc. The normalisation constant $\Sigma_0$ is consequently adjusted to account for a finite inner radius: 
\begin{equation}
\label{eq:surfacedensityring}
\begin{aligned}
\Sigma_0 &=
\frac{(2-n)M_{\rm disc} }{2\pi(r_{\rm out} ^{2}-r_{\rm in} ^{2}(\frac{r_{\rm out} }{r_{\rm in} })^n)},\: {\rm for} \: n\neq 2,\\
\Sigma_0 &=
\frac{M_{\rm disc} }{2\pi \ln({r_{\rm out} /r_{\rm in} }) r_{\rm out} ^2},\: {\rm for} \: n=2.
\end{aligned}
\end{equation}
which can be used for all values of $n$. For $n<2$ and $r_{\rm in} =0$ we recover Equation~\ref{eq:surfacedensity0schleicher}. Physically motivated values for $r_{\rm in}$ include the dust sublimation radius or $\sim$3 times the binary separation, corresponding to the innermost stable orbit in circumbinary systems \citep{Artymowicz1994}.

\subsubsection{Surface Density Profiles of Viscous Discs}
\label{sssec:surface_density_profiles_of_viscous_discs}

While power-law models are commonly used to describe disc surface density profiles due to their simplicity, they are not physically motivated and rely on poorly defined parameters such as the outer radius, $r_{\rm out}$. Observationally, different diagnostics yield inconsistent estimates of $r_{\rm out}$ - for instance, optically thin dust continuum observations often indicate smaller disc sizes than optically thick molecular line tracers \citep[e.g.,][]{2009Andrews}.

A more physically motivated model, widely adopted for YSO discs, is based on viscous disc evolution theory. This approach originates from the seminal works of \citet{1974lyndenbell} and \citet{1981pringle}, and was formulated in a practical surface density profile by \citet{1998hartmann}. In this model, the disc’s kinematic viscosity is assumed to follow a radial power law with index $\gamma$: $\nu \propto r^\gamma$. Under this assumption, the surface density profile takes the form:

\begin{equation}
\label{eq:surfacedensity_visc_discs}
    \Sigma(r) = \Sigma_0 \left( \frac{r}{r_{\rm ch}} \right)^{-\gamma} 
    \exp \left[ -\left( \frac{r}{r_{\rm ch}} \right)^{2 - \gamma} \right],
\end{equation}
where $r_{\rm ch}$ is the characteristic radius, and $\Sigma_0$ is the normalisation constant. Unlike the outer radius in power-law models, $r_{\rm ch}$ is well-defined within the viscous evolution framework and avoids the sensitivity associated with defining a sharp truncation.

Following \citet{Booth_2020}, the normalization constant $\Sigma_0$ for this profile can be expressed in terms of the total disc mass $M_{\rm disc}$ and $r_{\rm ch}$ as:

\begin{equation}
\label{eq:sigma0_viscous}
\Sigma_0 = (2 - \gamma) \frac{M_{\rm disc}}{2\pi r_{\rm ch}^2} \exp \left[ \left( \frac{r_{\rm in}}{r_{\rm ch}} \right)^{2 - \gamma} \right],
\end{equation}
where $r_{\rm in}$ is the inner radius of the disc, if the disc has one.

This exponentially tapered surface density profile has been successfully applied to fit the data from the observation of many YSO discs; however, it has not yet been widely applied to characterise post-AGB discs. 

\section{Gravitational Instability in YSO Discs}
\label{sec:ysos}
 Before applying our gravitational instability framework to the less explored environments of post-AGB discs, we first test its applicability using a set of well studied YSO systems. Many of these systems show direct or indirect features -- such as spirals, clumps, or large gaps -- that have been associated with gravitational instabilities \citep{Cadman2021}. By applying our analytical model to these systems, we aim to establish a benchmark for interpreting results in post-main sequence discs.

The YSO systems we consider are:
\begin{enumerate}
    \item AB Auriga: a system with a gravitationally unstable disc and a detected planet \citep{2022currie}.
    \item HL Tauri: a disc showing concentric rings with gaps, possibly indicating embedded planets \citep{2015ALMA,Booth_2020}.
    \item Elias 2-27: a system with spiral structures and density waves, suggesting possible gravitational instability \citep{Paneque_Carre_o_2021}.
    \item L1448 IRS3B: a triple protostar system embedded in a massive, gravitationally unstable disc with prominent spiral arms and a massive clump, as revealed by ALMA observations \citep{Reynolds_2021}.
    \item GQ Lupi: a T Tauri star with a $10-40$~M$_{\rm Jup}$ companion at $\sim110$~au. While the current disc mass is relatively low ($\sim$3–4~M$_{\rm Jup}$), the wide separation of the companion suggests it may have formed via gravitational instability during an earlier, more massive phase of the disc \citep{Wu_2017}. We note that the companion may lie in the brown dwarf regime, though its wide separation and possible disc origin make it relevant to gravitational instability-driven formation scenarios.
\end{enumerate}
For comparison, we also include two YSO systems with active planet formation, but without prominent indicators of disc instability:
\begin{enumerate}
    \item PDS 70: a low-mass disc system with two directly imaged planets \citep{Keppler_2019}.
    \item HD~169142: a system with a disc that shows gaps hinting at two planets, one of which has recently been confirmed through direct imaging and ALMA molecular line observations \citep{Fedele_2017, Hammond_2023}.
    \end{enumerate}

In Table \ref{tab:yso} we summarise the relevant properties of the YSOs used in our analysis, based on studies that have modelled the disc using observational data. For these systems, we have adopted the density profiles based on what was used in the respective studies, where some used the surface density profiles of viscous discs (Equation \ref{eq:surfacedensity_visc_discs}), while others used a power law surface density profile (Equation \ref{eq:surfacedensityring}; see Table \ref{tab:yso} and \ref{ap:appendix1} for more details). For HL Tauri and Elias 2-27, where stellar temperatures and radii are not available, we adopt plausible values consistent with their estimated stellar masses. We did not find an estimation for the inner radius of the disc of PDS 70 in the literature, hence we set its inner radius to $r_{\rm in}=0$.

\begin{figure}[hbt!]
\centering
\includegraphics[width=\linewidth]{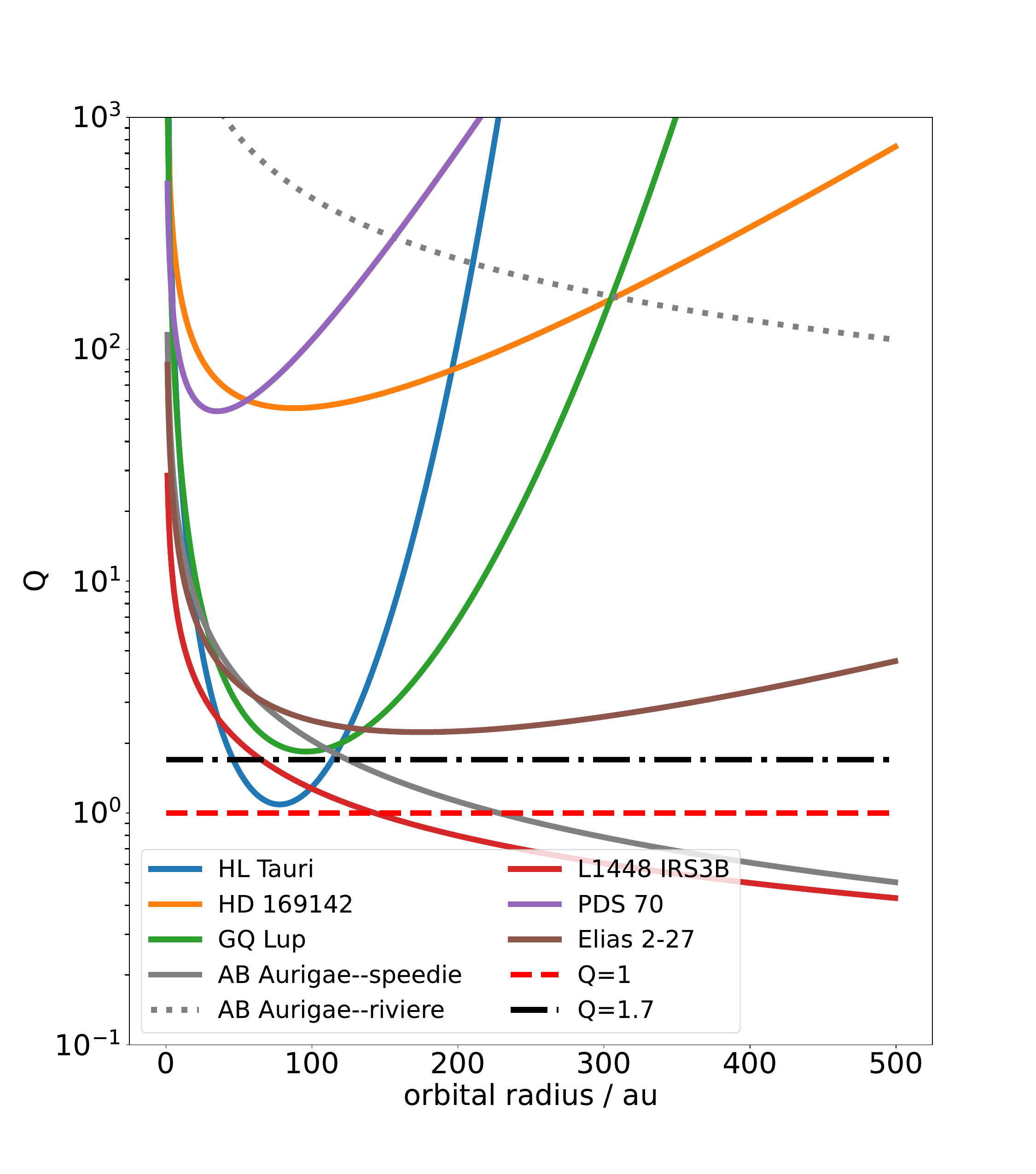}
\caption{Toomre parameter $Q$ versus disc radius for the YSOs tabulated in Table~\ref{tab:yso}. Only AB Aurigae with the disc mass reported in \citet{2023Speedie}, and L1448 IRS3B fall below $Q=1$.}
\label{fig:q_r_ysos}
\end{figure}

In Figure~\ref{fig:q_r_ysos}, we plot the Toomre parameter, $Q$ (see Section \ref{sec:cooling}), as a function of disc radius for each system. While our focus here is on the Toomre criterion as a first-order diagnostic, we note that gravitational fragmentation also requires rapid cooling (see Section~\ref{sec:cooling}), which we do not assess in this section. We find that only two of the systems are Toomre-unstable at present (AB Aurigae with the disc mass reported in \citet{2023Speedie}, and L1448 IRS3B) indicating that gravitational instability is not generally active under current disc conditions for most of the systems, although some of them such as HL Tauri, GQ Lup, and Elias 2-27 are close to the unstable regime. We note that Toomre analyses have been carried out in the past for HL Tauri \citep{Booth_2020} and Elias 2-27 \citep{Perez_2016}, and the resulting Toomre profiles in these studies match the profiles we found for them. As a further benchmark of our method, we compute the Toomre Q profiles of three protoplanetary discs with published Q profiles, as detailed in \ref{ap:benchmark}.

Our findings suggest that planet formation via gravitational instabilities may have occurred in these YSOs earlier in the disc's evolution. We revisit this possibility in Section~\ref{sec:earlylife}, where we explore whether these systems could have been gravitationally unstable at earlier stages. 

\begin{table*}[hbt!]
\begin{threeparttable}
\caption{Stellar and disc parameters for the benchmark YSO systems used in our gravitational instability analysis (see \ref{ap:appendix1} for sources).}
\label{tab:yso}
\resizebox{\columnwidth}{!}{
\begin{tabular}{lccccccc}
\toprule
\headrow System & AB Aurigae & HL Tauri & Elias 2-27 & PDS 70 & 
HD 169142 & L1448 IRS3B & GQ Lup \\

\midrule
Planets detected & 1 & none\footnote{No planets confirmed via direct imaging or RV, though disc structures suggestive of embedded planets exist \citep{2015ALMA}} & none & 2&1 & none\footnote{The mass of the clump observed in L1448 IRS3B  is high enough to be a protostar.}  & 1 \\
 disc mass, $M_{\rm disc}$ ($\rm M_\odot$)& $0.0032$, $0.7$\footnote{Two values exist in the literature; see \ref{ap:appendix1}; in this analysis, we have plotted the Toomre curve of this disc using both disc mass values in Figure \ref{fig:q_r_ysos}.} & $0.2 $ &$0.08 $ &$0.003 $ & $0.008$  & $0.29$ &$0.0733$ \\
Inner radius, $r_{\rm in}$ (au)& $80$  & $8.78$ & $5$  & $0$  \footnote{No value available - we set the inner radius to 0.}  & $13$ &$0.1$ & $1.7$\\ 
Characteristic/Outer radius\footnote{For all systems except AB Aurigae and L1448 IRS3B, the tabulated value in this row is the characteristic radius. For AB Aurigae and L1448 IRS3B, the tabulated values are their outer radii.}, $r_{\rm ch}$/$r_{\rm in}$ (au)& $500$&$ 80$ & $200$ & $40$& $100$  & $299$&$19.5$\\
Star mass, $M_{\rm star}$ ($\rm M_\odot$)& $2.4$&$1.7 $&$0.46 $&$ 0.85$& $1.65$ &$1.2$ &$1.05$\\
Star temperature, $T_{\rm star}$ (K) &$9770$& $4000$&$3850$ &$3972$ & $7500$ &$5000$\footnote{Indirectly calculated from the luminosity of the protostar, assuming a blackbody spectrum.} &$4300$\\
Star radius, $R_{\rm star}$ ($\rm R_\odot$)& $2.5$& $6.9 $ & $ 2.3$& $1.26$& $1.6$ & $2.5$&$1.7$\\
$\gamma$ & -- & $-0.2$ & $1$& $1$& $1$ &-- &$-0.21$\\
$n$ & $1$ & -- & --& --&-- &$1.2$ &--\\
\bottomrule
\end{tabular}}   
\end{threeparttable}
\end{table*}


\section{Gravitational Instability in Typical Post-AGB Discs}
\label{sec:postagbs}

Having used our gravitational instability framework on well-studied YSO systems, we now apply the same methodology to the post-AGB binary population. 

Estimates of post-AGB disc gas masses, based on ALMA observations of molecular line emission (e.g., CO), typically fall in the range of $10^{-4}$–$10^{-2}$~M$_{\odot}$ \citep[e.g.][]{Bujarrabal_2013, Gallardo_Cava_2021}. In contrast, disc dust masses inferred from radiative transfer modelling of near and mid-infrared interferometric data and spectral energy distributions (SEDs) imply significantly higher total disc masses - reaching up to $0.1$–$0.2$~M$_\odot$ in a few individual cases \citep[e.g.,][]{Kluska_2018, Corporaal_2023}. These estimates are sensitive to assumptions about grain properties, vertical structure, and disc inclination, and are often subject to model degeneracies. Interpreting such high dust masses with a canonical gas-to-dust ratio of $\sim$100 \footnote{A key uncertainty in interpreting both gas and dust measurements is the gas-to-dust ratio in evolved systems. While a ratio of $\sim$100 is commonly assumed in protoplanetary discs, post-AGB discs may have undergone significant chemical evolution and/or gas depletion. The true ratio remains unconstrained, and thus dust-based masses may overestimate, and gas-based masses may underestimate, the total disc mass.}
 would imply total disc masses of order $10^{-1}$~M$_{\odot}$, in tension with dynamical constraints and gas-based mass estimates \citep{Hillen_2015,Kluska_2018}. To account for the uncertainties in total disc mass estimates - arising from the discrepancy between gas and dust measurements and the unknown gas-to-dust ratio - we adopt a conservative range of $0.01$–$0.1$~M$_\odot$ for the total (gas + dust) disc mass. We assess gravitational stability across this range using a power-law surface density profile, consistent with standard characterisations of post-AGB discs from observations (Section~\ref{ssec:the_surface_density_profile}). Typical truncation radii lie between 100 and 500~au; for our fiducial case, we adopt $r_{\rm out} = 100$~au to represent the more compact end of the distribution, where surface densities are highest and instability most likely.



We assume a central binary mass of $M_1+M_2=1.5$~M$_\odot$, consistent with typical post-AGB binaries \citep[e.g.,][]{Oomen_2018}, and adopt $R_{\rm star} = 100$~R$_\odot$ and $T_{\rm star} = 5,000$~K for the post-AGB star. Assuming a binary separation of $1$~au, we set the disc inner radius to the expected dynamical truncation radius of $r_{\rm in}=3$~au following \citep{Artymowicz1994}. 

\begin{figure*}[hbt!]
\centering
\includegraphics[width=\linewidth]{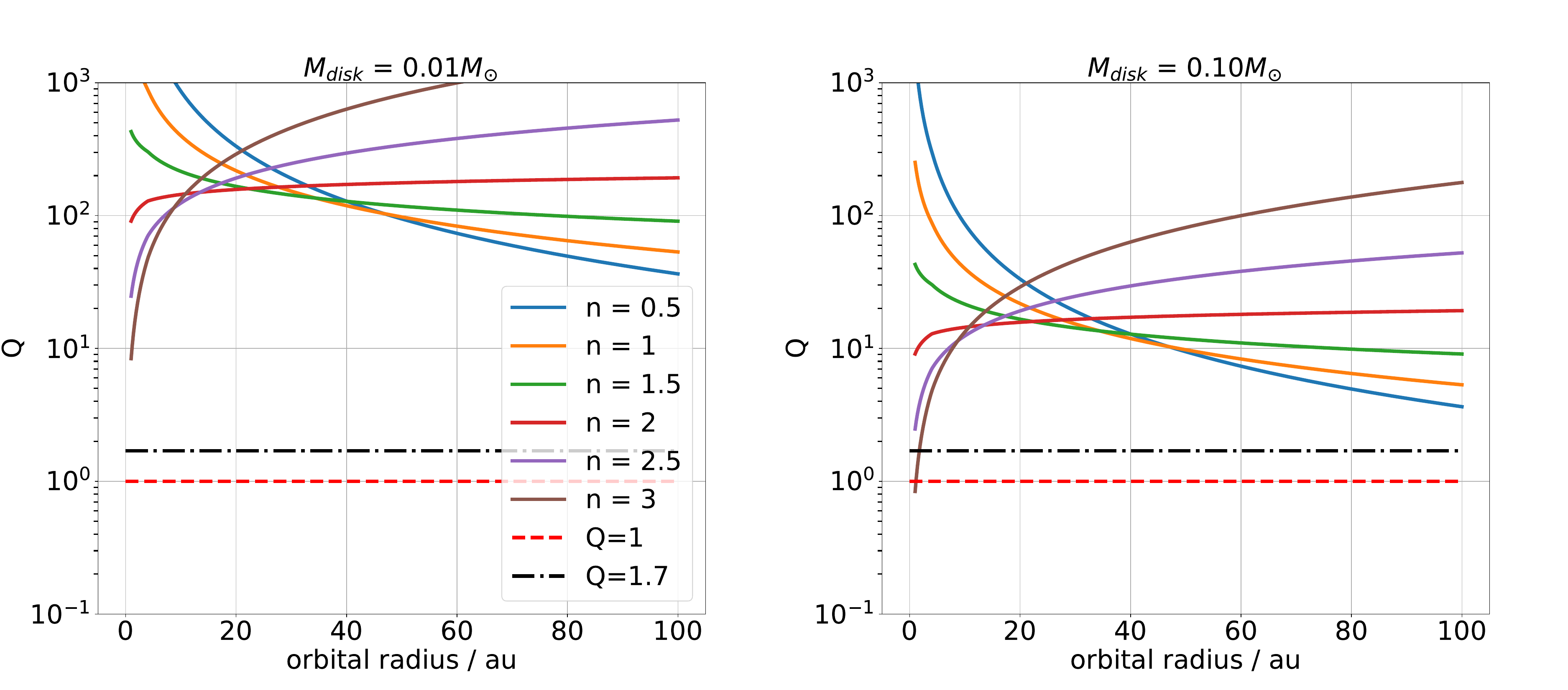}
\caption{The Toomre parameter $Q$ as a function of radius for typical post-AGB discs, assuming $M_1+M_2=1.5$~M$_\odot$, $R_{\rm star} =100$~R$_\odot$, $T_{\rm star} =5,000$~K, $r_{\rm in}=3$~au, and $r_{\rm out}=100$~au as a lower limit of post-AGB disc truncation radii. The index $n$ is the power-law index of the surface density profile for each of the curves. Even in this relatively compact scenario, we see that the discs are generally stable.}
\label{fig:toomrepagb}
\end{figure*}

In Figure~\ref{fig:toomrepagb}, we show the Toomre parameter $Q$ as a function of radius for discs with varying surface density power-law indices. Post-AGB discs appear generally stable across this range. For power-law indices $n\geq 2$, the surface density declines steeply with radius, and stability increases accordingly due to the rapidly decreasing mass at large radii. However, at smaller radii, discs with steeper profiles ($n>2$) can approach marginal instability.  

To reflect the observed structure and ensure physical consistency, we adopt a power-law (Equation \ref{eq:surface_density_power_law}) to parametrise the surface densities of post-AGB discs following how they are typically modelled in literature, and we also include an inner cavity in our analysis (Section~\ref{ssec:the_surface_density_profile}, Equation \ref{eq:surfacedensityring}) to avoid this divergence and to remain consistent with the typical observed structure of post-AGB discs. This is in contrast with how we have parametrised YSO discs (Equation \ref{eq:surfacedensity_visc_discs}) in the preceding section.

Even under compact, high-density configurations, the resulting $Q$ values remain above the fragmentation threshold in most regions, indicating that typical post-AGB discs are gravitationally stable under present-day conditions.

\subsection{Testing the Stability of the Massive Disc in IRAS 08544-4431}
\label{sec:iras}

IRAS~08544-4431 is a post-AGB system with an infrared excess, indicating the presence of an optically thick circumbinary disc. Its inner rim lies at the dust sublimation radius \citep{Corporaal_2023}, consistent with a disc in an early evolutionary stage. \footnote{The inner rim of the {\it dust} disc observed in the infrared may not coincide with the inner radius of the {\it gas} disc, which may extend further inward.} 

For stellar parameters, we adopt $T_{\rm star}=7250$~K, $R_{\rm star} = 65$~R$_\odot$, $M_1=0.75$~\msun{}, and $M_2=1.65$~M$_\odot$.  For the disc, we adopt an outer radius of $r_{\rm out} =175$~au and the inner radius of three times the binary orbital separation ($a=1.74$~au), giving $r_{\rm in} =5.22$~au \citep{Kluska_2018}. 

We explore a plausible range of disc masses: at the high end we adopt $M_{\rm disc} = 0.1$~M$_\odot$ which we adopt from modelling carried out by \citet{Corporaal_2023} based on their observations of the dust component, who found a dust mass of $M_{\rm disc} = 0.001$~M$_\odot$ under the assumption of a gas-to-dust ratio of 100\footnote{\citet{Kluska_2018} previously found a dust mass of $0.002$~M$\odot$ and used a gas-to-dust ratio of 100, giving $M_{\rm disc}=0.2$~M$\odot$. However, \citet{Corporaal_2023} provide more recent and accurate measurements, which we adopt here.}. At the lower end, we adopt the gas mass estimated by \citet{Bujarrabal_2018} from ALMA CO Observations: $M_{\rm disc}=0.02$~M$_\odot$.

\citet{Corporaal_2023} modelled the surface density using a piecewise power law, with a break at $r_{\rm mid}=3~R_{\rm in}$, and power-law indices of $n_{\rm in}=1.5$ and $n_{\rm out}=-1$ in the inner and outer regions, respectively: 

\begin{equation}
\label{eq:sigmapiecewise}
\begin{aligned}
\Sigma(r) = \Sigma_0\left(\frac{r_{\rm mid}}{r}\right)^{n_{\rm in}}
\, \text{for} \, r_{\rm in}<r<r_{\rm mid} ,\\
\Sigma(r) = \Sigma_0\left(\frac{r_{\rm mid}}{r}\right)^{n_{\rm out}}
\, \text{for} \, r_{\rm out}>r>r_{\rm mid} .
\end{aligned}
\end{equation}

The surface density normalisation, $\Sigma_0$, is then obtained from the total disc mass using (not valid for $n=2$):
\begin{equation}
\label{eq:sigm0piecewise}
    M_{\rm disc}=
\Sigma_0 \left( \frac{2\pi r_{\rm mid} ^{n_{\rm in}}(r_{\rm mid} ^{2-n_{\rm in}}-r_{\rm in} ^{2-n_{\rm in}})}{(2-n_{\rm in}) }+\frac{2\pi r_{\rm mid} ^{n_{\rm out}}(r_{\rm out} ^{2-n_{\rm out}}-r_{\rm mid} ^{2-n_{\rm out}})}{(2-n_{\rm out}) }\right)
\end{equation}

\begin{figure}[hbt!]
\centering
\includegraphics[width=8cm]{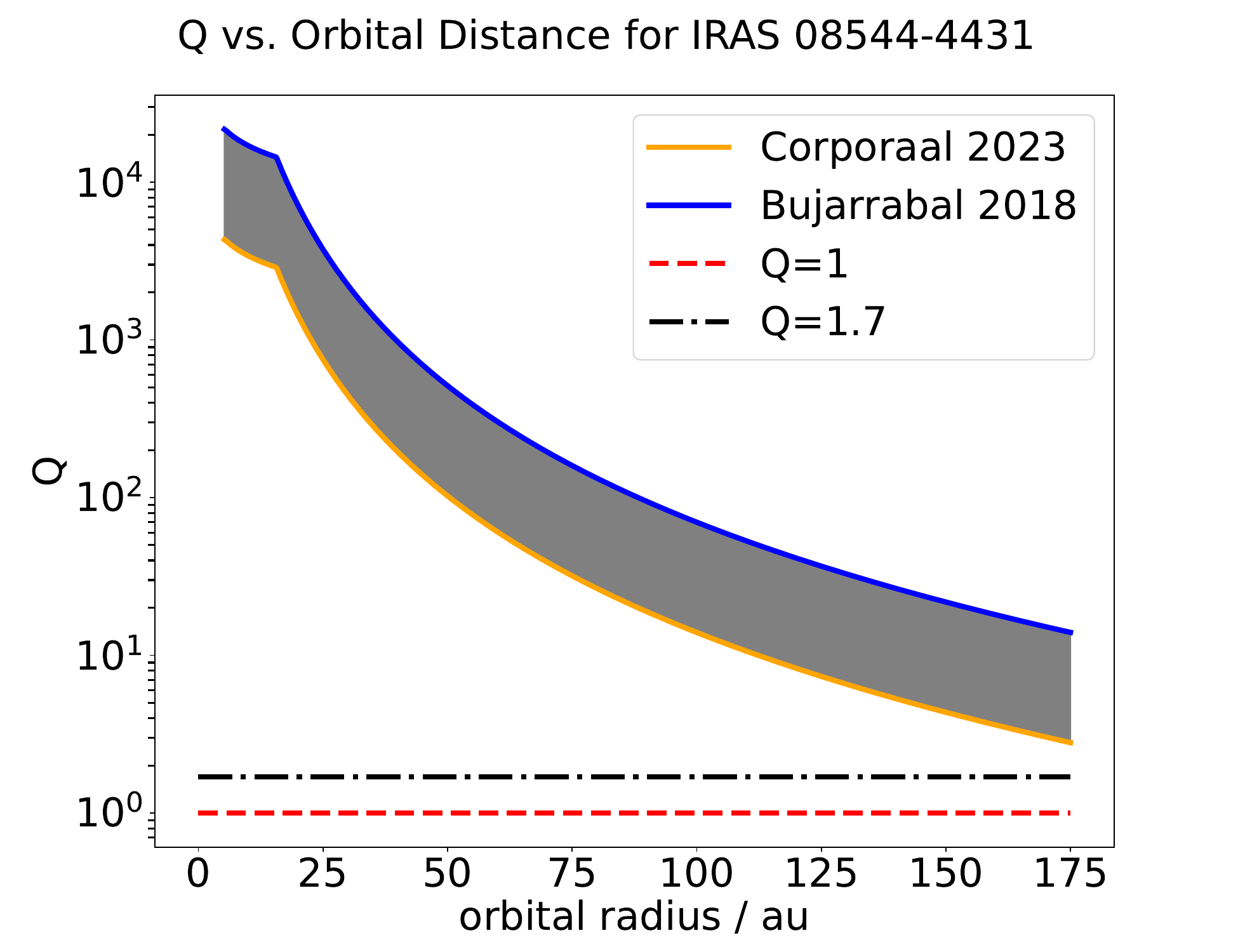}
\caption{The Toomre parameter $Q$ as a function of radius for the post-AGB system IRAS 08544-4431. The shaded region  indicates the uncertainty in the total disc mass, based on estimates from \citet{Bujarrabal_2018} and \citet{Corporaal_2023}.}
\label{fig:qvsriras}
\end{figure}

Figure~\ref{fig:qvsriras} shows the resulting Toomre profiles for both disc mass estimates. If the higher mass inferred by \citet{Corporaal_2023} is accurate, the disc reaches $Q < 4$ in its outer regions - approaching but not exceeding the fragmentation threshold. This places IRAS~08544$-$4431 among the most gravitationally unstable post-AGB discs in our sample, though it remains above the threshold for fragmentation. By contrast, the lower mass estimate from \citet{Bujarrabal_2018} yields a Toomre profile that stays well within the stable regime across all radii. These results highlight the sensitivity of the Toomre parameter to disc mass assumptions and underscore the need for more robust constraints on gas and dust masses in evolved systems.

\subsection{An Exploration of the Parameter Space for Gravitational Instability Across Different Disc Masses and Sizes}
\label{sec:grid_pagb}

To complete our investigation, we explore the gravitational stability of discs across a range of mass and size configurations. Specifically, we plot the Toomre parameter $Q$ as a function of disc mass and outer radius to assess whether a given system - such as a post-AGB binary with a more massive disc in the past - could have supported planet formation via gravitational instability. 

\begin{figure*}[hbt!]
\centering
\includegraphics[width=6cm]{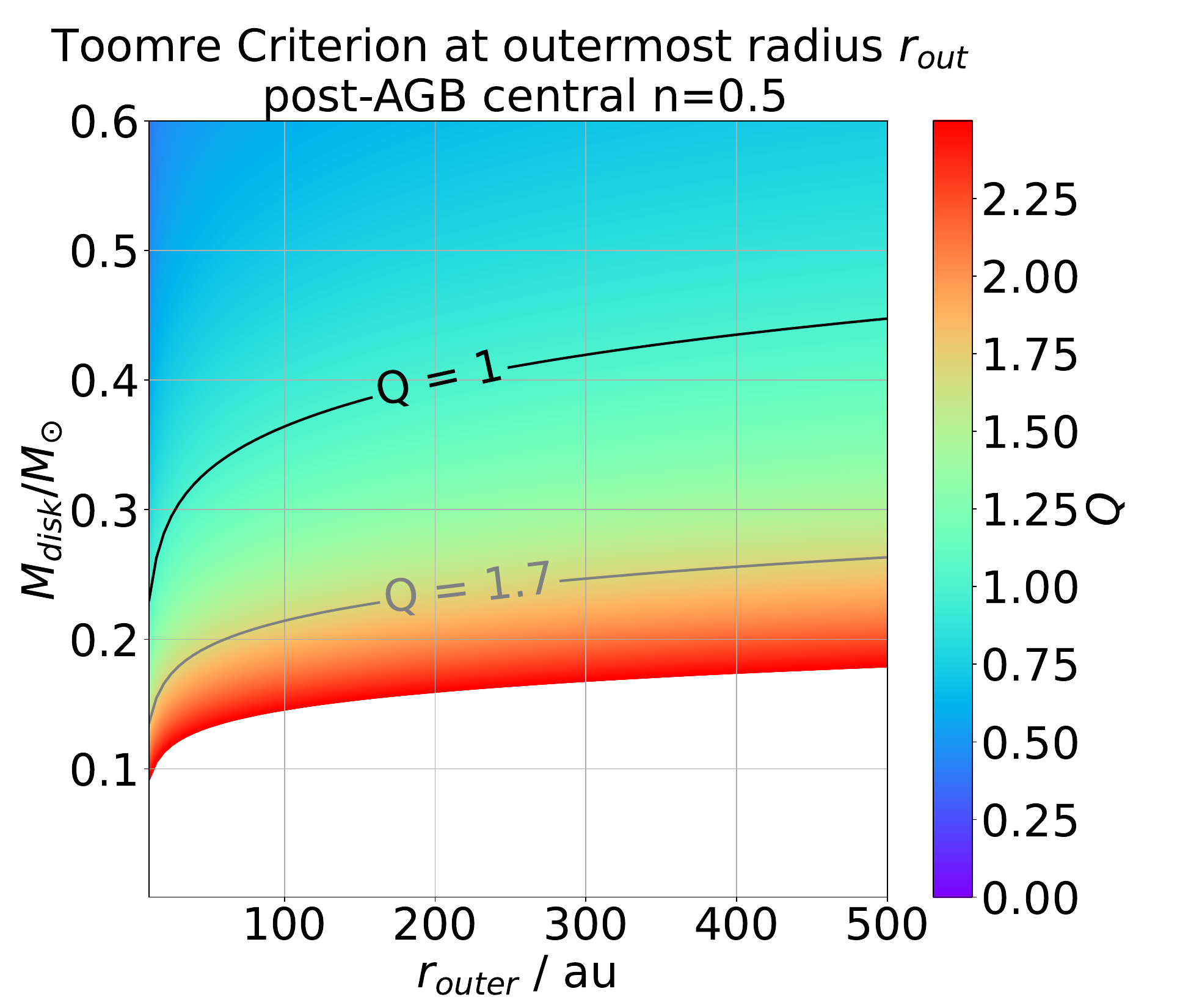}
\includegraphics[width=6cm]{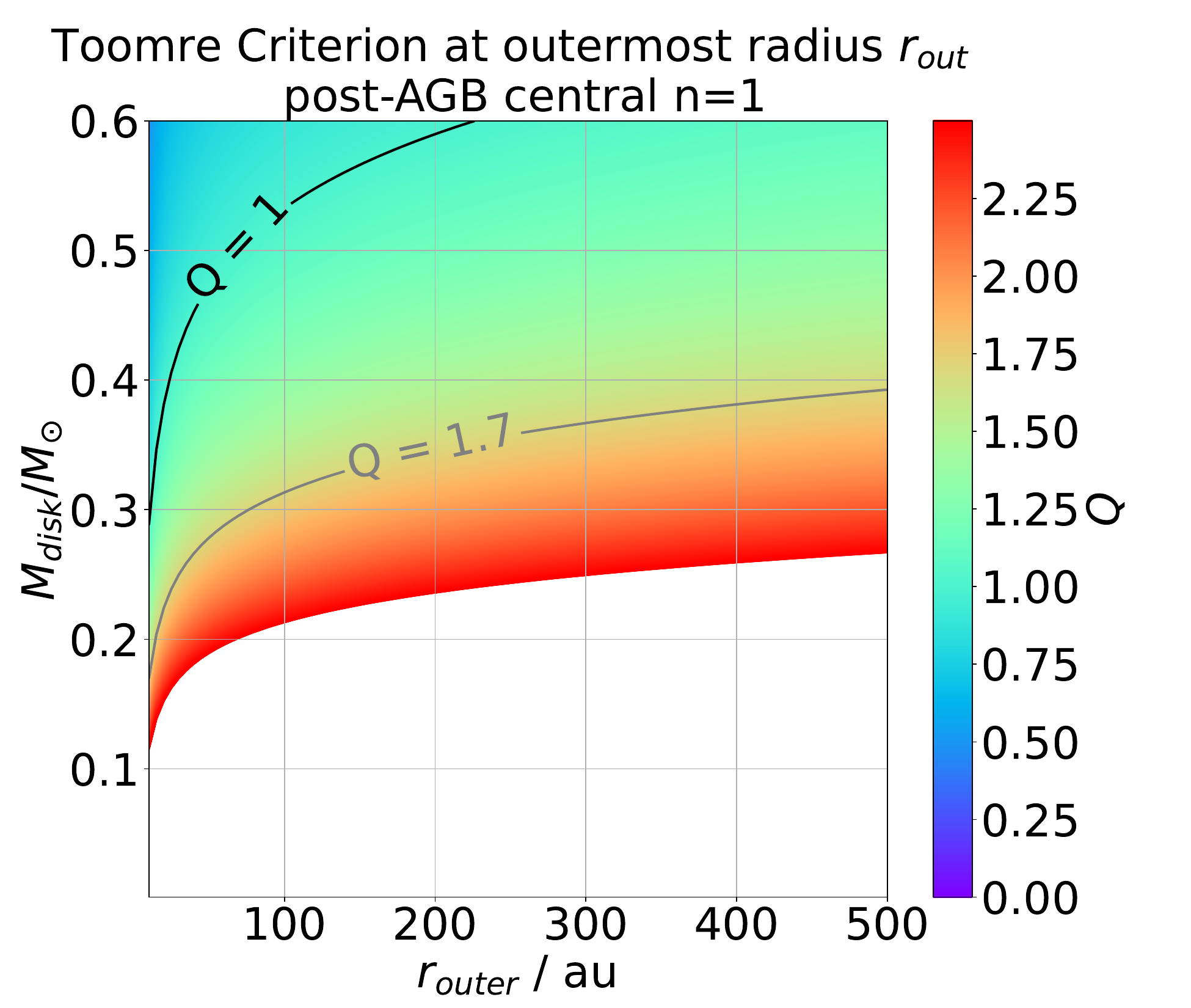}
\includegraphics[width=6cm]{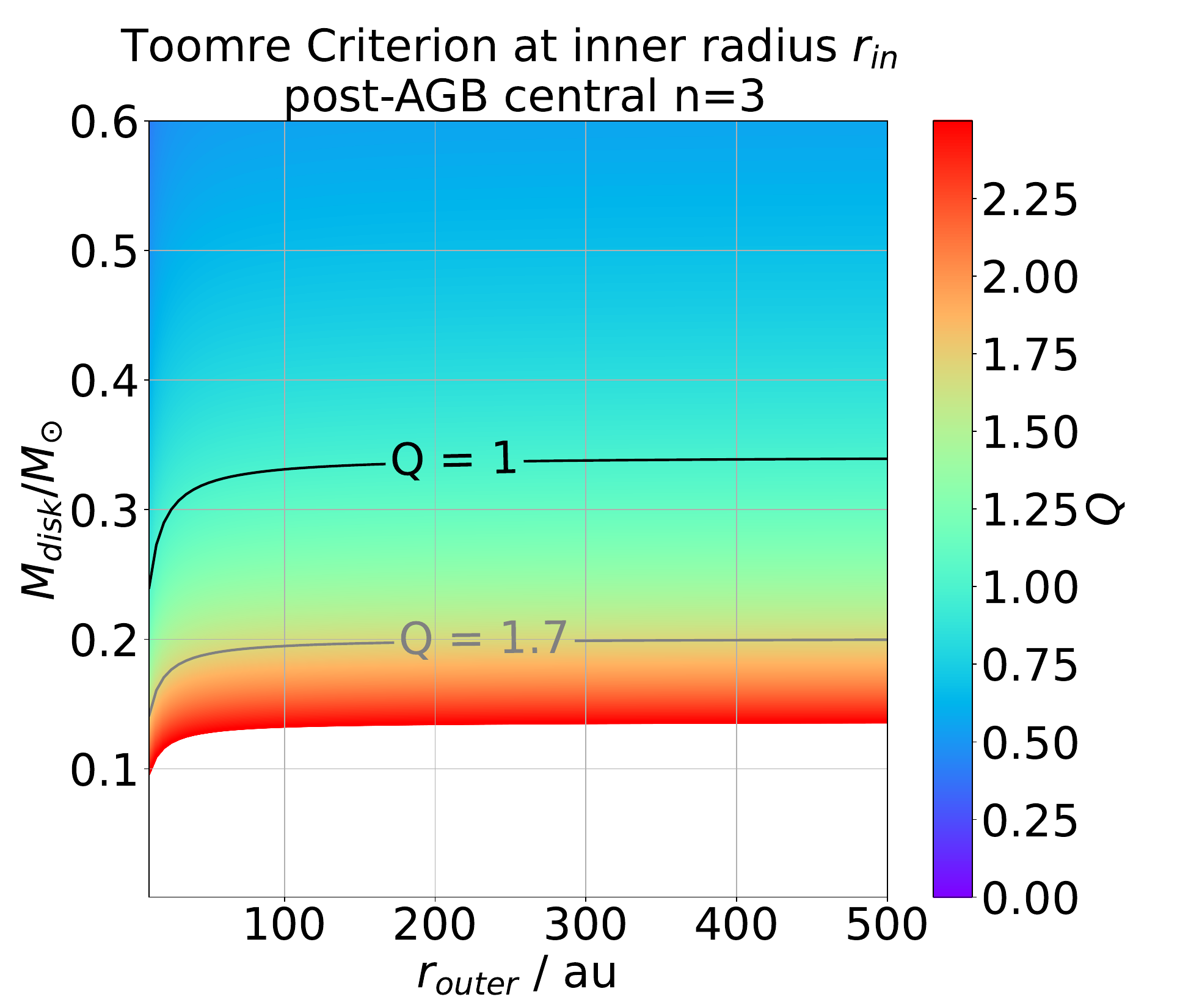}
\caption{Contours showing variation of the Toomre parameter $Q$ as a function of disc radius and mass, evaluated at the outermost radius $r_{\rm out}$ for post-AGB discs with surface density exponent $n=0.5$ and $n=1$ (left and centre panels), and at the innermost radius $r_{\rm in}$ for $n=3$ (right panel).
All calculations adopt $R_{\rm star}=100$~R$_\odot$, $T{\rm star}=5000$~K, and $M_{\rm cnt}=1.5$~M$_\odot$.}
\label{fig:grid_pagb}
\end{figure*}

\begin{figure}[hbt!]
\centering
\includegraphics[width=9cm]{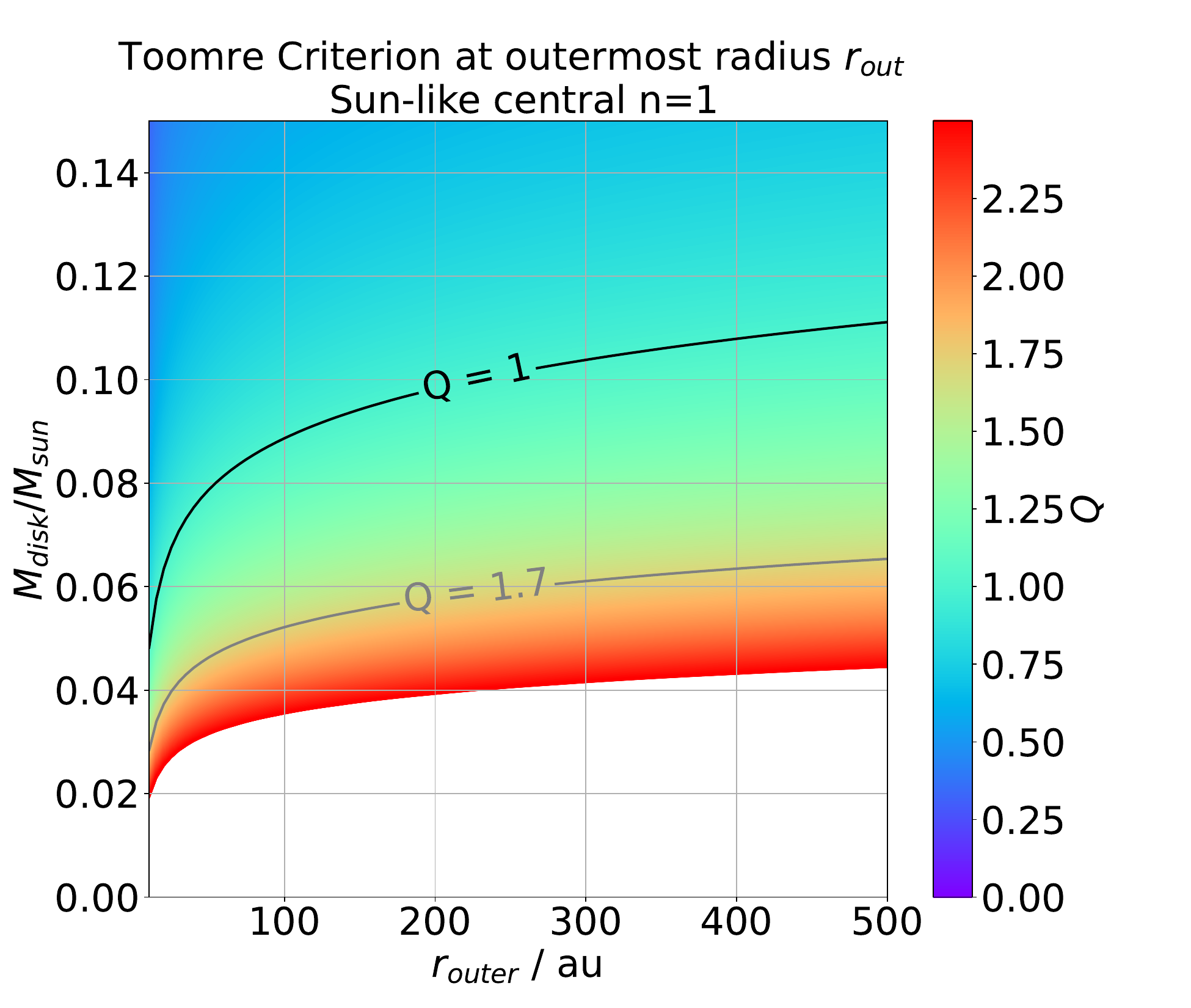}
\includegraphics[width=9cm]{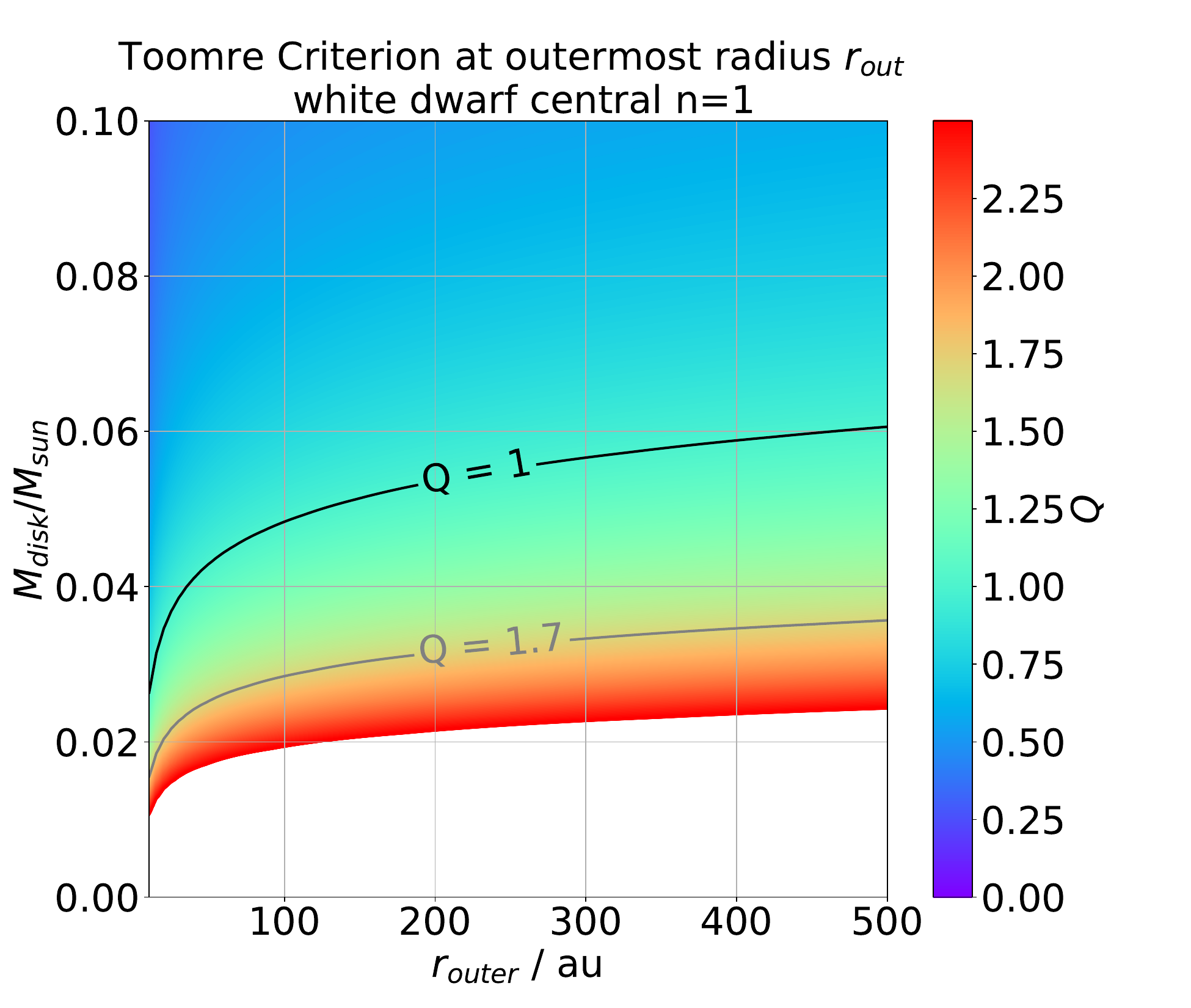}
\caption{Toomre parameter $Q$ as a function of disc mass and outer radius, evaluated at $r = r_{\rm out}$, for discs around a Sun-like star (top) and a white dwarf (bottom), assuming a surface density exponent $n = 1$. Fixed parameters are $R_{\rm star}=1$R$\odot$, $T{\rm star}=6000$~K, $M_{\rm cnt}=1$M$_\odot$ (top figure for the sun-like star), and $R_{\rm star} =0.02$R$_\odot$, $T_{\rm star} =30,000$~K, $M_{\rm cnt}=0.5$M$_\odot$ (bottom figure for the white dwarf). }
\label{fig:grid_discmassouterrad_sun}
\end{figure}

We consider three central star cases: post-AGB binaries (Figure~\ref{fig:grid_pagb}), main-sequence stars (Figure~\ref{fig:grid_discmassouterrad_sun}, top panel), and white dwarfs (Figure~\ref{fig:grid_discmassouterrad_sun}, bottom panel). These plots provide a reference framework for gauging the gravitational stability of discs around stars at different evolutionary stages.

In all cases, we adopt a power-law surface density profile (Equation~\ref{eq:surface_density_power_law}) with $n=1$, and additionally show $n=0.5$ for post-AGB systems. For steep profiles ($n=3$), where $Q$ decreases with radius, we evaluate $Q$ at the inner radius ($r_{\rm in}$) instead of $r_{\rm out}$, as shown in the right panel of Figure~\ref{fig:grid_pagb}. We adopt fixed parameters of  $r_{\rm in} =3$~au (corresponding to the dynamical truncation radius for a post-AGB binary system with an orbital separation of $a=1$~au).
The contours represent the value of $Q$ at the disc's outer edge ($r = r_{\rm out}$), which corresponds to the minimum $Q$ for surface density slopes $n < 2$. For $n > 2$, where $Q$ decreases outward, the minimum occurs at the inner edge, and we report that value instead. 

We find that only sufficiently massive and compact discs are prone to enter the Toomre-unstable regime. Discs with representative post-AGB disc masses ($\sim0.01$~M$_\odot$) are unlikely to becomes unstable under present-day conditions and support planet formation via gravitational instability. This is in contrast with discs around main-sequence stars and white dwarfs (Figure \ref{fig:grid_discmassouterrad_sun}) which can become gravitationally unstable with much lower disc masses.

We next examine whether the likely conditions of post-AGB discs at the time of their formation would have made them more susceptible to instability and whether planet formation could have occurred during that earlier phase.

\section{Were Discs Less Stable Earlier in Their Lifetimes?}
\label{sec:earlylife}

More massive and radially compact discs are more prone to gravitational instability. If the discs we observe today were more massive and compact in the past, they may have been unstable at earlier epochs -- even if they appear stable today. Here we use viscous evolution theory of discs to determine their properties in the past and to determine whether, at that time, they may have been gravitationally unstable.

Viscous spreading naturally causes discs to expand and redistribute mass over time \citep{1974lyndenbell,1981pringle,1998hartmann}. The evolution of the disc's surface density under viscous processes is governed by the diffusion equation: 
\begin{equation}
\label{eq:diff_visc_disc}
    \frac{\partial \Sigma}{\partial t} = \frac{3}{r} \frac{\partial}{\partial r} \left[ r^{1/2} \frac{\partial}{\partial r} \left( r^{1/2} \nu \Sigma \right) \right]
\end{equation}
where $\nu$ is the kinematic viscosity. Assuming $\nu$ varies with radius as a power law, $\nu \propto r^\gamma$ (index is $\gamma$), the radial evolution of the surface density can be solved analytically \citep{1998hartmann}:
\begin{equation}
\label{eq:surfacedensity_hartmann}
    \Sigma(r,T)=\frac{C}{3\pi \nu_0} \left( \frac{r}{r_{\rm ch,0}}\right)^{-\gamma}T ^{-\frac{2.5-\gamma}{2-\gamma}} \exp \left (-\left( \frac{r}{r_{\rm ch,0}}\right)^{2-\gamma}\frac{1}{T}\right)
\end{equation}
where $C$ is a scaling constant, $\nu_0=\nu(r=r_{\rm ch,0})$, $r_{\rm ch, 0}$ is the characteristic radius at $t=0$, and $T=1+t/t_{\rm visc}$ is a dimensionless time. The viscous timescale of the disc, $t_{\rm visc}$, is given by:
\begin{equation}
\label{eq:t_visc}
    t_{\rm visc}=\frac{r_{\rm ch, 0}^2}{3(2-\gamma)^2\nu_0}.
\end{equation}
Equation~\ref{eq:surfacedensity_hartmann} is the time-dependent analogue of Equation~\ref{eq:surfacedensity_visc_discs} from Section~\ref{sssec:surface_density_profiles_of_viscous_discs}, showing that surface density profiles spread out from initially more centrally concentrated configurations. The characteristic radius evolves as \citep{2009Andrews}
\begin{equation}
    \label{eq:radius_viscous_spread}
    r_{\rm ch}(T)=r_{\rm ch, 0} T ^{\frac{1}{2-\gamma}}
\end{equation}

Under the same assumptions, the total disc mass decreases over time due to accretion. \citet{2009Andrews} give the mass evolution as:
\begin{equation}
    \label{eq:mass_viscous_spread}
    M_{\rm disc}(T)=M_{\rm disc, 0} T ^{\frac{-1}{2(2-\gamma)}},
\end{equation}
where $M_{\rm disc, 0}$ is the initial disc mass. In principle, Equations~\ref{eq:radius_viscous_spread} and \ref{eq:mass_viscous_spread} can be used to infer past disc properties if the age is known. However, disc ages are poorly constrained for both YSOs and post-AGB systems. Stellar ages cannot be used as reliable proxies for disc ages, particularly in YSOs \citep{2009Andrews}. 
 
Here, we assume that post-AGB disc evolution is also dominated by viscous processes, as in YSO discs, although the influence of binary companions may modify this picture. The gravitational torques from companions could affect the disc's evolution, making these equations less valid \citep{1998hartmann}. However, since fragmentation is typically expected in outer disc regions, we assume binary effects are second-order for this analysis.

It is unclear how long discs - even in the case of YSOs - remain in the viscous evolution regime, i.e., how far back in time the analytical relations of Equations~\ref{eq:radius_viscous_spread} and \ref{eq:mass_viscous_spread} apply. However, the uncertainty in the viscosity parameter and disc age can be mitigated by expressing the evolution of the disc’s mass and radius in terms of the number of viscous timescales that have elapsed, where each viscous timescale depends on the initial radius via Equation~\ref{eq:t_visc}.

\begin{figure*}[hbt!]
\centering
\includegraphics[width=\linewidth]{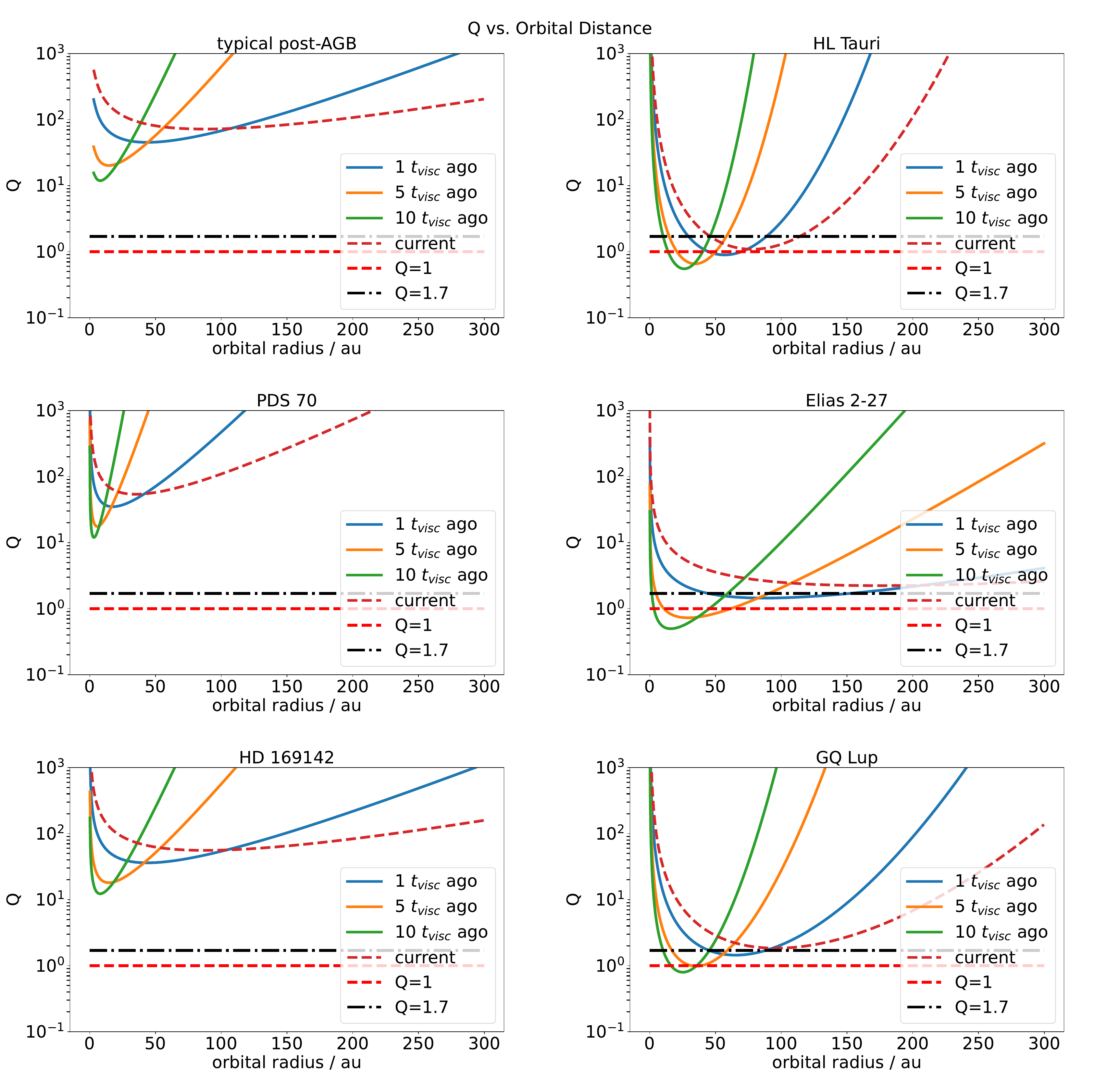}
\caption{Toomre parameter $Q$ as a function of orbital radius for our YSO sample (excluding AB Aurigae and L1448 IRS3B, for which a characteristic radius is not available) and a representative post-AGB disc (central object parameters similar to those used in Figure~\ref{fig:toomrepagb}; a typical value of $\gamma=1$ was assumed due to lack of empirical data). We show $Q$ evaluated at the initial state of each disc for the different assumed viscous timescales.}
\label{fig:earlyysos}
\end{figure*}

In Figure \ref{fig:earlyysos}, we plot the Toomre parameter $Q$ as a function of radius for several YSO discs and a representative post-AGB disc 
($M_{\rm disc, now}=0.02$~M$_\odot$, $r_{\rm ch, now}=100$~au, and assuming $\gamma=1$), evolved backwards in time for different numbers of elapsed viscous timescales  ($T=t/t_{\rm visc}+1$). We assume constant inner radii during viscous evolution. For post-AGB discs, the inner radius is set by the dynamical truncation radius, while for YSOs, it is set by the sublimation radius (taken here as the radius where $T_{\rm subl}=1200$~K). These boundaries depend on stellar properties rather than viscous evolution. Figure \ref{fig:earlyysos} shows that even at earlier epochs, some of the discs were likely gravitationally stable - particularly in the cases of PDS 70, HD169142, and the post-AGB example. However, in the case of GQ Lup, HL Tau, and Elias 2-27, they could have been gravitationally unstable in their pasts.

Our analysis includes all YSO systems discussed in Section~\ref{sec:ysos}, except AB Aurigae and L1448 IRS3B, whose previously estimated density profiles are not compatible with the form in Equation~\ref{eq:surfacedensity_visc_discs}. For post-AGB discs, we assume they follow the viscously evolving form of Equation~\ref{eq:surfacedensity_visc_discs}, thus assigning them a characteristic radius. This differs from the simpler power-law model used in earlier sections. 

To convert viscous timescales into ages, we adopt the Shakura–Sunyaev \citep{Shakura1973} \(\alpha\)-disc prescription for viscosity ($\nu$):

\begin{equation}
    \label{eq:viscosity}
    \nu=\alpha c_s H 
\end{equation}
where $\alpha$ is the viscosity efficiency parameter, and $c_s$ and $H$ are the sound speed and scale height, respectively. The scale height $H$ of the disc can be found with the following equation: 
\begin{equation}
\label{eq:scale_height}
    H(r)=\frac{c_{\rm s}(r)}{\Omega(r)}.
\end{equation}

Typical values of $\alpha$ in planetary discs range from 0.0005 to 0.08, based on observed accretion rates \citep{2009Andrews}. Combining this with the expressions for $c_s$ and $H$ (Equations~\ref{eq:sound_speed} and \ref{eq:scale_height}) and setting $\eta=1$, the radial dependence of viscosity becomes: 

\begin{equation}
    \nu(r) =\alpha \left (\frac{k_{\rm B}T_{\rm sub}}{2m_{\rm p}}\right ) \left (\frac{R_{\rm sub}^{0.75}}{(GM_{\rm cnt})^{0.5}}\right )r^{0.75}
\end{equation}

Outside the sublimation radius, we evaluate $\nu$ at $r = r_{\rm ch, 0}$ to determine $\nu_0$. Here, we assume that the stellar parameters have not evolved significantly over the relevant timescales, so the sublimation radius remains approximately constant. For convenience, we group all constants into a single term $K$, allowing us to write:
\begin{equation}
    \nu_0 =Kr_{\rm ch, 0}^{0.75},
\end{equation}
yielding a viscous timescale of (for $\gamma=1$): 
\begin{equation}
\label{eq:visc_t2}
    t_{\rm visc}=\frac{r_{\rm ch, 0}^{1.25}}{3K}
\end{equation}
Thus, the disc age can be expressed as $t_{\rm visc}\times(T-1)$. 

To calculate these parameters, we adopt  representative post-AGB values similar to those used in Section \ref{sec:postagbs} and Figure \ref{fig:toomrepagb}. Disc ages for various assumed viscosity parameters ($\alpha$) are presented in Table \ref{tab:disc_ages}. Importantly, varying $T=t/t_{\rm visc}+1$ alters both the inferred disc age and the initial characteristic radius $r_{\rm ch, 0}$, but we find that the total system age remains relatively stable for each specific value of $\alpha$.

We find that the disc age does not vary significantly with changes in the scaled time $T$ for a given $\alpha$, since the viscous timescale itself adjusts accordingly. However, for fixed $T$, the disc age is inversely proportional to $\alpha$ for $\gamma=1$. The stellar ages compiled from the literature are included for comparison and should be considered upper limits to the corresponding disc ages. In general, lower $\alpha$ values yield disc ages that exceed the stellar ages - consistent with the findings of \citet{1998hartmann}, who argued that $\alpha \approx 0.01$ is typical for YSO discs. For Elias 2-27 specifically, \citet{2009Andrews} derived $\alpha = 0.03$ from accretion rate measurements, which supports our finding that an assumed $\alpha = 0.01$ overestimates the disc age relative to the stellar age.

For post-AGB systems, the inferred ages are consistent with their expected lifetimes of $\sim$$10^5$~yr \citep{Bujarrabal_2016} for the higher end of viscosity parameters we adopted; $\alpha=0.01$.

\begin{table*}[hbt!]
\begin{threeparttable}

\caption{Disc age parameters estimated for each system by assuming different values of the viscosity parameter $\alpha$. For each case, we list the corresponding viscous timescale ($t_{\rm visc}$), number of viscous timescales elapsed ($t/t_{\rm visc}$), inferred disc age ($t$), and the initial characteristic radius ($r_{\rm ch,0}$). For post-AGB discs, we assume $\gamma = 1$ due to a lack of empirical values; for YSOs, $\gamma$ is taken from Table~\ref{tab:yso}. See Section ~\ref{sec:earlylife} for more details.}

\label{tab:disc_ages}
\resizebox{\columnwidth}{!}{
\begin{tabular}{lccccc}
\toprule
\headrow System &  viscosity parameter & viscous timescale & elapsed viscous timescales   & disc age & initial characteristic radius  \\
-&  $\alpha$ &$t_{\rm visc}$& $t/t_{\rm visc}$& $t$& $r_{\rm ch,0}$\\
-& - &(yr) &-&(yr)&(au)\\
\midrule
typical post-AGB &$10^{-2}$ & $5.4\times10^4$& $1$ & $5.4\times10^4$ & $50$ \\
 &$10^{-2}$ & $1.4\times10^4$& $5$ & $6.9\times10^4$& $17$\\
 & $10^{-2}$ & $6.5\times10^3$& $10$ & $6.5\times10^4$& $9.1$\\
 & $10^{-3}$&$5.4\times10^5$ & $1$ & $5.4\times10^5$ & $50$\\
 & $10^{-4}$&$5.4\times10^6$ & $1$ & $5.4\times10^6$ &  $50$\\
\midrule
HL Tauri &$10^{-2}$ & $7.3\times10^5$& $1$ & $7.3\times10^5$ & $58$\\
 &$10^{-2}$ & $3.9\times10^5$& $5$ & $2.0\times10^6$ & $35$\\
 &$10^{-2}$& $2.8\times10^5$& $10$ & $2.8\times10^6$& $27$ \\
 & $10^{-3}$& $7.3\times10^6$& $1$ & $7.3\times10^6$&  $58$\\
 &  $10^{-4}$& $7.3\times10^7$ & $1$ &  $7.3\times10^7$& $58$ \\
 \midrule
Elias 2-27 &$10^{-2}$& $1.8\times10^6$& $1$ & $1.8\times10^6$& $100$\\
& $10^{-3}$& $1.8\times10^7$ & $1$ & $1.8\times10^7$ & $100$ \\
&  $10^{-4}$& $1.8\times10^8$ & $1$ & $1.8\times10^8$ & $100$ \\
 \midrule
PDS 70 & $10^{-2}$ & $4.9\times10^5$& $1$ & $4.9\times10^5$& $20$\\
 & $10^{-3}$&$4.9\times10^6$ & $1$ & $4.9\times10^6$ & $20$\\
 &  $10^{-4}$& $4.9\times10^7$ & $1$ & $4.9\times10^7$& $20$\\
 \midrule
HD 169142 &$10^{-2}$& $6.9\times10^5$& $1$ &  $6.9\times10^5$& $50$\\
& $10^{-3}$& $6.9\times10^6$ & $1$ & $6.9\times10^6$& $50$\\
&  $10^{-4}$& $6.9\times10^7$ & $1$ &  $6.9\times10^7$ & $50$\\
 \midrule
GQ Lupi &$10^{-2}$ & $1.8\times10^6$& $1$ &  $1.8\times10^6$ & $68$\\
&$10^{-3}$& $1.8\times10^7$ & $1$ & $1.8\times10^7$& $68$\\
&  $10^{-4}$& $1.8\times10^8$ & $1$ & $1.8\times10^8$& $68$ \\
\bottomrule
\end{tabular}}   
\begin{tablenotes}[hang]
\item[] The estimated ages for the central stars of these discs are 1~Myr for HL Tauri \citep{Booth_2020}, 1~Myr for Elias 2-27 \citep{2009Andrews}, 5.4~Myr for PDS 70 \citep{Keppler_2019}, 6~Myr for HD~169142\citep{Hammond_2023}, and 2-5~Myr for GQ Lupi \citep{Wu_2017}.
\end{tablenotes}
\end{threeparttable}
\end{table*}

\subsection{The Effects of Varying the Temperature Profile}

In the analysis so far, we have assumed that the disc temperature follows the radial dependence given by  Equation \ref{eq:Tprofile}. However, the temperature gradient is linked to the viscous disc's surface density index $\gamma$ in Equation~\ref{eq:surfacedensity_hartmann}. Specifically, if $T(r)\propto r^p$, and $\nu\propto r^\gamma$, then $\gamma=1.5+p$ \citep{1998hartmann}. 

In reality, the assumption of constant $\alpha$ may not hold, and the Shakura-Sunyaev prescription may not strictly apply. Therefore, we have not varied the temperature profile as a function of the observationally derived $\gamma$ values in our main analysis. However, in this section we explore how changes in the temperature index $p$ affect the gravitational stability of a post-AGB disc, using the corresponding value of $\gamma=1.5+p$ in each case. 

We test values of $p=0.5,0.75,1$, motivated by the constraints from \citet{Kluska_2019} on the radial temperature structure in these systems. This analysis is presented assuming physically motivated viscous surface density profiles (Equation \ref{eq:surfacedensity_visc_discs}) for post-AGB discs. In contrast, for the rest of the study, post-AGB discs are modelled with observationally inferred power-law surface densities, where no such $p$–$\gamma$ coupling exists.

\begin{figure}[hbt!]
\centering
\includegraphics[width=\linewidth]{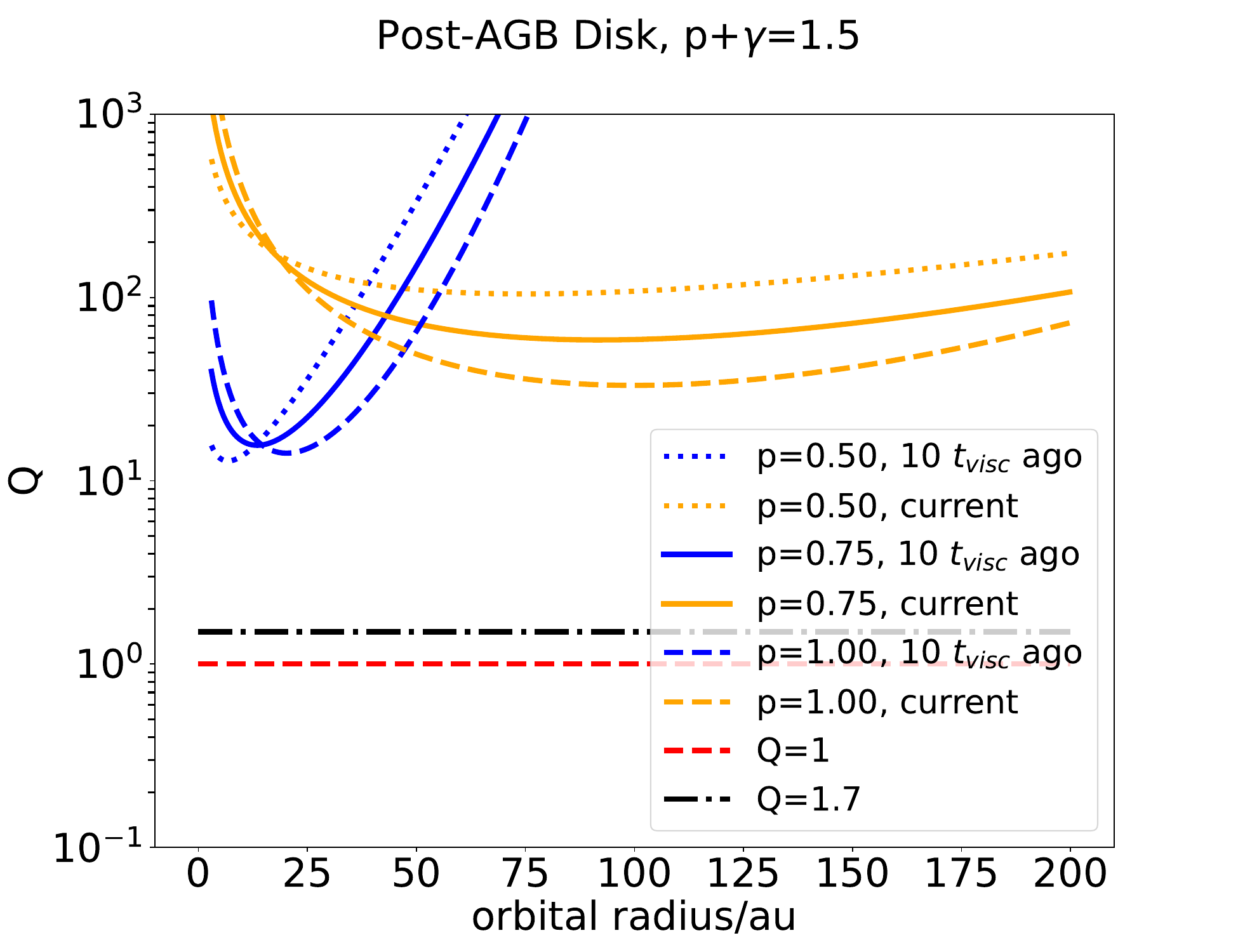}
\caption{The Toomre parameter $Q$ as a function of orbital radius for a representative post-AGB disc for varying temperature profile index, $p$, and the corresponding $\gamma$, assuming $\gamma=1.5+p$. The central binary properties are similar to Figure \ref{fig:toomrepagb}, while assuming the surface density profile of a viscous disc (Equation \ref{eq:surfacedensity_visc_discs}). We adopt $r_{\rm ch}=100$~au and an inner dynamic truncation radius of $r_{\rm in}=3$
~au. }
\label{fig:earlypagb_varyt}
\end{figure}

Figure \ref{fig:earlypagb_varyt} shows the resulting $Q$ profiles. As expected, steeper temperature profiles (larger $p$) lower the $Q$ parameter across the disc. However, even for the steepest case considered, the disc remains gravitationally stable. We emphasise that this variation does not include different temperature slopes across the sublimation radius, as applied in earlier figures.


\section{Planet Formation in Post-Common Envelope Discs}
\label{sec:nnser}

In this section we apply our gravitational instability framework to the post common envelope binary NN~Ser, where the presence of two proposed circumbinary planets has been interpreted as evidence for second generation planet formation in a remnant disc. These planets were first inferred from eclipse timing variations \citep{Beuermann_2010}. Given their proximity to the central binary, it has been argued that they could not have formed during the system's main sequence evolution and subsequently survived both the giant phase and the common envelope expansion. Instead, it has been proposed that they formed within a circumbinary disc created during the common envelope interaction.

The existence of these circumbinary planets, however, has been the subject of debate. While the planets were first inferred from eclipse timing variations \citep{Beuermann_2010, Beuermann_2013}, and later studies concluded the proposed orbits to be dynamically stable \citep[][]{2012Horner,Marsh_2013}, the interpretation of the observations as being due to planets is nonetheless regarded with caution. In particular, \citet{Marsh_2013} noted that the planets are only indirectly detected, and that alternative mechanisms -- most notably magnetic activity cycles (Applegate’s mechanism \citep{1992Applegate}) -- can also reproduce the observed eclipse timing variations.

On the assumption of the planets’ existence, \citet{2014schleicher} developed an analytical framework to estimate the properties of a (non observed) circumbinary disc formed by a common envelope ejection.  Then  applying gravitational instability theory, they demonstrated that the planetary masses and orbital separations inferred for NN Ser are consistent with planet formation in such an environment. Here we connect our gravitational instability framework applied to (observed) post-AGB discs to their analysis of NN Ser, and in so doing we re-examine several aspects of their model.

By considering common envelope physics, \citet{2014schleicher} derived a disc mass of $M_{\rm disk}=0.146$~M$_\odot$\footnote{It is not clear what post-CE disc masses may be, with simulations indicating the ability of the interaction to eject almost the entire envelope leaving very little mass to form a disc \citep[e.g.,][]{2015ivanova,2020Reichardt,Gonzalez2022}. Observationally, only very low mass discs have been detected around some post-CE binaries \citep{Li2025}.}. With this disc mass, they went on to {\it assume} that the entire disc around NN~Ser is marginally gravitationally unstable ($Q(r)\sim 1$) in order to reproduce the observed planet masses. 

Here we use our framework to determine whether their disc would be unstable to planet formation. We use their disc mass, their surface density (a power-law index of $n=1$ for the surface density profile -- Equation~\ref{eq:surface_density_power_law}), and their central binary mass, $M_{\rm core}+M_2=0.646$~M$_\odot$. However, we use a more realistic temperature profile than used by them, in line with our analysis - Equation \ref{eq:Tprofile}, where the disc is heated by irradiation from the central source, which in the case of NN~Ser is a white dwarf. Finally, we use a disc outer radius of $r_{\rm out} =7.50$~au, calculated by conservation of angular momentum before and after disc formation - see \ref{ap:nnser} for details.

\begin{figure}[hbt!]
\centering
\includegraphics[width=\linewidth]{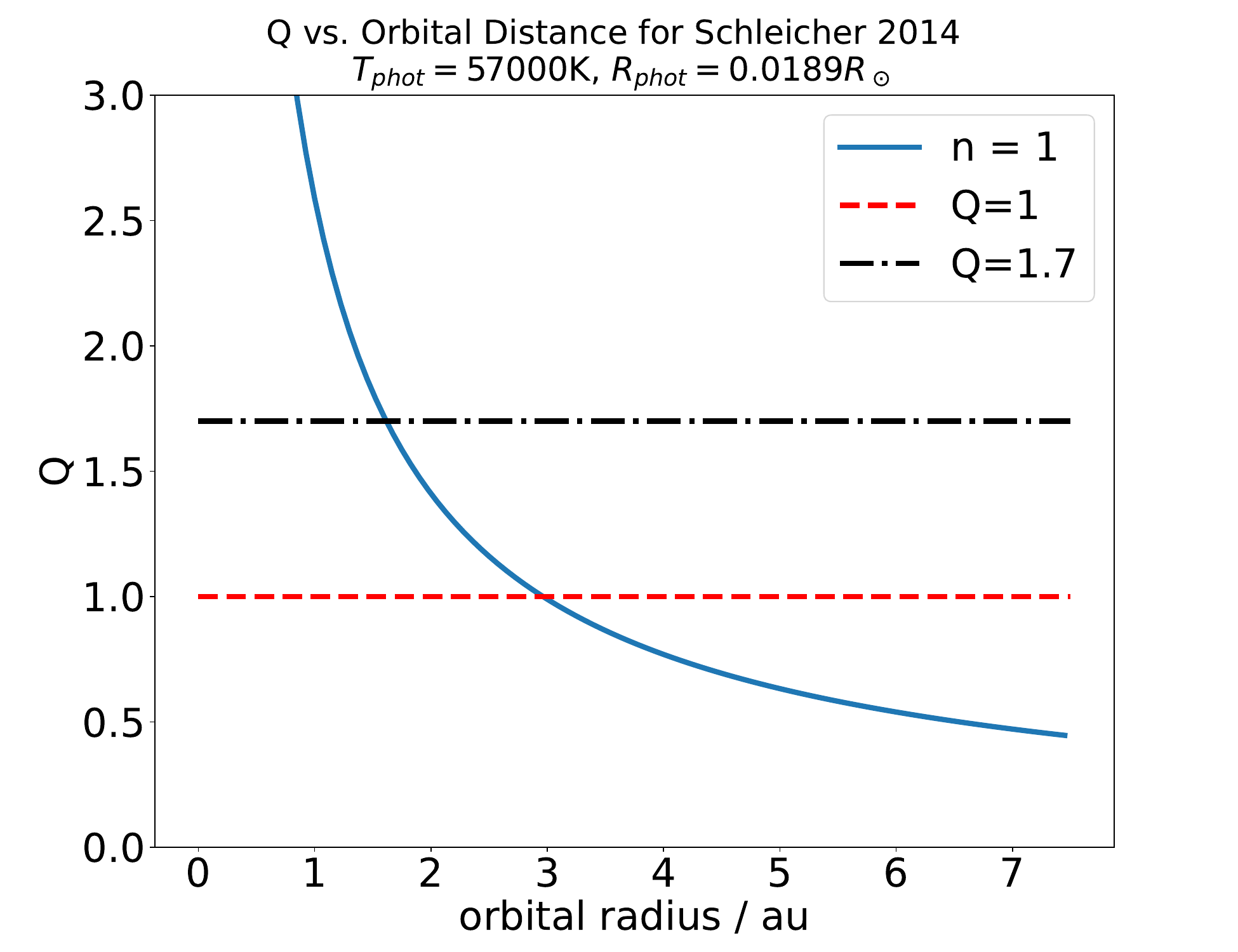}
\caption{The Toomre parameter, $Q$, as a function of orbital radius for the progenitor disc of NN~Ser, based on the values of \citet{2014schleicher}, which are summarised in Table \ref{tab:schleicher_nnser}. The white dwarf parameters were $T_{\rm eff} = 57,000$~K and $R_{\rm star} =0.0189$~R$_\odot$, and the mass of the central binary was set to be $M_{\rm c}+M_2=0.646$~M$_\odot$. We note that the outer radius of the disc used to reproduce this plot was $r_{\rm out} =7.50$~au, which we recomputed using the value we find for angular momentum in \ref{ap:nnser}.}

\label{fig:Qvsr_Tradial}
\end{figure}

Figure~\ref{fig:Qvsr_Tradial} shows the radial variation of the Toomre parameter, $Q$, for NN~Ser. The figure shows that even with a more realistic thermal structure, the disc becomes gravitationally unstable beyond $\sim3$ au - precisely where the proposed planets are located. This analysis therefore confirms that the gravitational instability mechanism remains viable under more physically motivated conditions than those selected by \citet{2014schleicher}, but only if the disc mass is as high as they propose.

\section{Conclusion}

In this study, we have carried out a comprehensive analytical investigation into whether post-AGB circumbinary discs can support planet formation via gravitational instability. Our key findings are: 

\begin{enumerate}
    \item Under realistic assumptions of disc mass $\sim0.02$~M$_\odot$, disc structure, and density and temperature profiles, post-AGB discs are generally gravitationally stable at present. Even systems with the most optimistic parameters - such as IRAS 08544-4431 with a disc mass of $0.1$M$_\odot$ - remain above the fragmentation threshold, with Toomre $Q>1$ throughout the disc, largely due to thermal support from the luminous central star. In contrast, discs around lower-luminosity central objects (e.g., white dwarfs or Sun-like stars) can reach the gravitationally unstable regime at lower masses.
    \item Using self-similar viscous disc evolution models, in which the surface density profile retains its shape while the overall scale (mass and characteristic radius) evolves due to viscous spreading, we explored whether post-AGB discs may have been unstable earlier in their lifetimes. We found that, even after evolving them backward in time across several viscous timescales, most remain stable, unlike some protoplanetary discs (e.g., HL Tauri, Elias 2-27, GQ Lupi), which likely passed through an unstable phase during earlier epochs. Our analysis also shows that higher viscosity parameters ($\alpha \sim 10^{-2}$) are more consistent with expected post-AGB disc lifetimes.
    \item Revisiting the post-common envelope binary NN Ser, we confirm that if the disc parameters proposed by  \citet{2014schleicher} are assumed, the disc would become gravitationally unstable beyond $\sim\,3$~au. However recent developments in common envelope theory suggest that fallback discs may be far less massive than previously assumed.

\end{enumerate}

It is important to note that using the Toomre parameter as a diagnostic for gravitational instability, while useful, does not take factors such as vertical structure, turbulence, and binary-driven dynamics into account. Post-AGB discs can differ from YSO discs in these regards as well, hence the results of this study should be treated as a first order analysis. 

Overall, our analysis demonstrates that gravitational instability is highly unlikely to account for second generation planet formation in typical post-AGB discs, as fragmentation would require unrealistically high disc masses ($\gtrsim 0.4$ M$_\odot$), even under compact and cool conditions. The observed signatures of planets or planet-like companions in several post-AGB systems therefore point towards alternative formation channels - most likely a modified version of core accretion operating on shorter timescales in evolved environments. This study provides a first theoretical baseline for interpreting such systems. While future observations will better constrain disc masses, gas-to-dust ratios, temperatures, sizes, and lifetimes - and may uncover additional evidence for planet formation - we will, in parallel, explore additional theoretical models to investigate planet formation pathways in post-main-sequence environments.

\paragraph{Acknowledgments}
AP acknowledges the financial support provided by the International Macquarie Research Excellence Scholarship (iMQRES) program,
received throughout the duration of this research. DK, ODM
and MW acknowledge the support from the Australian Research Council Discovery Project DP240101150. We thank Toon De Prins for carefully reading the manuscript and pointing out several typographical errors. We also thank the anonymous referee for valuable comments that improved the paper.

\printendnotes

\bibliography{refs}

\appendix

\section{Parameters of YSO systems}
\label{ap:appendix1}
In this section we present the studies we have used to compile the YSO disc properties in Table~\ref{tab:yso}. 

We note that the reported inner radii in the literature typically refer to the inner radius of the dust component. Some studies explicitly report an inner gas cavity (such as HD169142), but others do not distinguish between the two. If a study does not state whether the reported inner radius  is the dust or gas radius, we have taken it to be the {\it gas} inner radius. For studies that carried out simulations and  reported their adopted inner radius, such as for AB Aurigae, or Elias 2-27, we have used their value. For PDS 70 there were no reported inner cavities in the literature, hence we adopted $r_{\rm in}=0$. 

There are two different reported masses for the disc of AB Aurigae, each obtained with different methods, and we have carried out Toomre analysis with both values. \citet{rivier2024} reported $3.2\times10^{-3}~\rm M_\odot$ which is based on ALMA and VLA observations to find the dust mass and then used a gas-to-dust ratio to estimate the total disc mass. On the other hand, \citet{2023Speedie} have estimated the disc mass based on their observation and subsequent modelling of the gravitational wiggle they observe in the disc, which they argued is a signature of gravitational instability, and found that a disc mass of $0.7~\rm M_{\odot}$ best fits their model. Other than the disc mass, the rest of the tabulated values are those that \citet{2023Speedie} have taken for their hydrodynamics simulation (note that AB Aurigae and L1448 IRS3B are the only systems which were modelled using a power law for the surface density rather than Equation~\ref{eq:surfacedensity_visc_discs}). 

For HL Tauri, the parameters for the star's radius and luminosity are from \citet{1999menshchikov}. The star's mass was calculated by \citet{2016pinte} by fitting models to CO and HCO$^+$ velocities obtained from ALMA, and the rest of the parameters, most notably the disc mass, were taken from \citet{Booth_2020}, who have used $^{13}\rm C^{17}\rm O$ measurements from ALMA to find a disc mass that is significantly more massive compared to previous measurements.

For Elias 2-27 , we adopt the value of the inner radius of the disc from \citet{Paneque_Carre_o_2021}, the photospheric temperature and radius of the star from \citet{2009Andrews} and the rest from \citet{Veronesi_2021}, who used ALMA observations to find the disc and star masses. It is noteworthy to mention that this system has been also analysed for unstable behaviour by \citet{Perez_2016}, albeit with different values, and for their standard opacity value their resulting Toomre estimates are similar to ours, noting the slightly different values they have adopted for the disc parameters. 

The values for PDS 70 are available in \citet{Keppler_2019}, from ALMA measurements. We assume the inner radius of the disc to be approximately zero, $r_{\rm in} = 0$~au.

For HD 169142, we take the values of star mass, radius, and temperature from \citet{Quanz_2013}. \citet{Quanz_2013} reported three different stellar surface temperatures, and the adopted one was calculated from high resolution spectroscopy by \citet{2006guimaraes}, who found $T_{\rm eff}=7500$~K. We take the characteristic disc radius and other disc parameters from \citet{Fedele_2017}, who used a best fit to ALMA observations to estimate the disc parameters. For the disc mass, \citet{Fedele_2017} mentioned that the total dust mass is $10^{-4}$~M$_\odot$ and the best fit they find for the gas-to-dust mass ratio is 80, which results in a  disc mass of 0.008~M$_\odot$. They also determined  $\Sigma_{\rm ch}=6.5$~g/cm$^2$ to be the surface density at the characteristic radius. 

For GQ Lup, both disc and star parameters have been summarised by \citet{Wu_2017}. The disc surface density was fitted using ALMA data. Similarly for L1448 IRS3B, the disc and star parameters have been taken from \citet{Reynolds_2021}, who used ALMA data.

\section{Benchmarking Against Published Toomre Q Profiles}
\label{ap:benchmark}
\begin{figure}[hbt!]
\centering
\includegraphics[width=\linewidth]{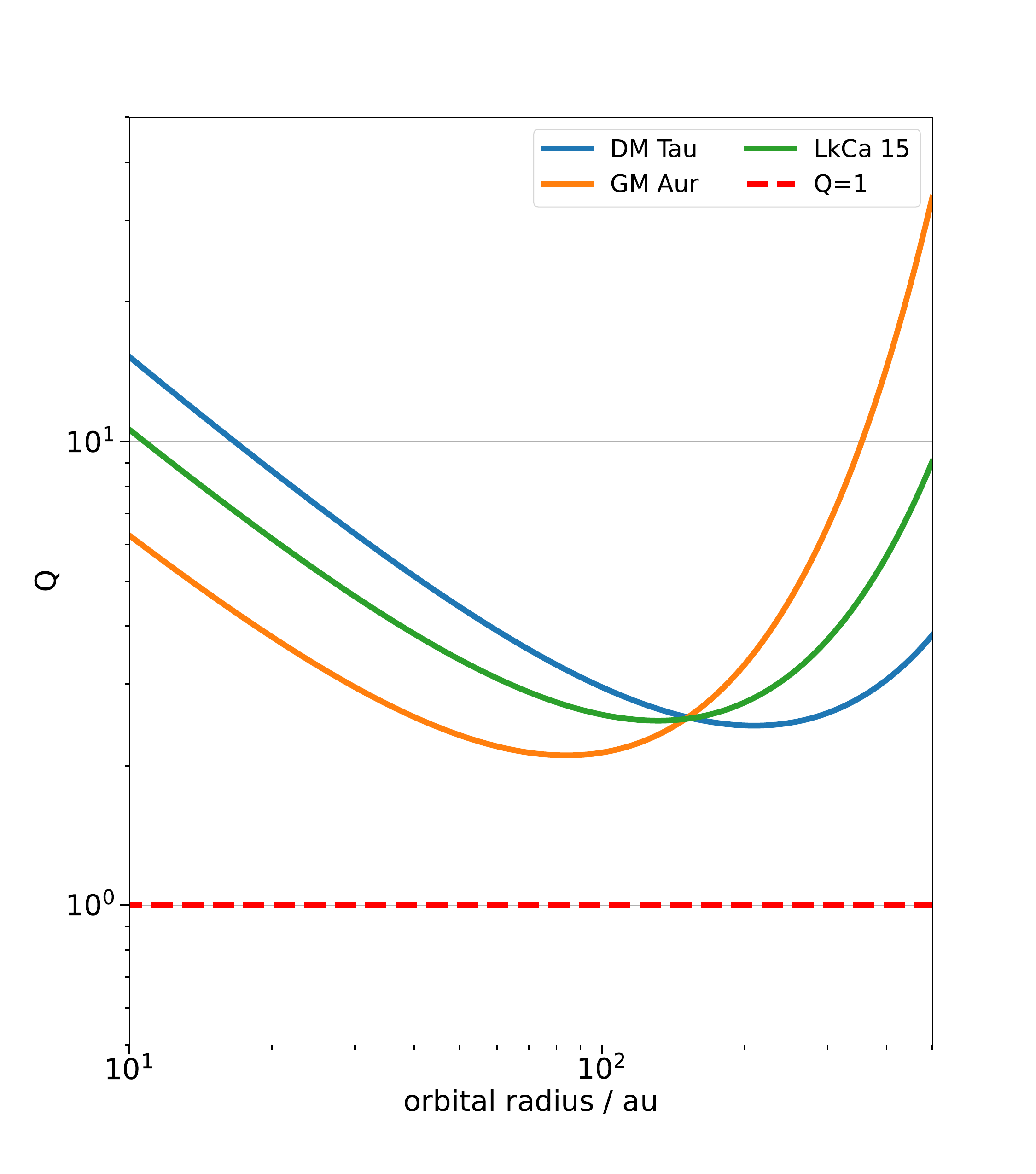}
\caption{Toomre parameter $Q$ versus disc radius for the YSO discs of DM Tau, GM Aur, and LkCa 15. The mass, and characteristic radius of these systems were taken from \citet{Martire_2024,Longarini_2025}, whilst their star temperature and radius were taken from \citet{2011Andrews}. We used the surface density profile of viscously evolving discs (Equation \ref{eq:surfacedensity_visc_discs}) to model these systems, assuming no inner cavities for them. See Figure 7 of \citet{Longarini_2025}, and Figure 9 of \citet{Martire_2024} for comparison.}
\label{fig:q_r_longarini}
\end{figure}

To further test the robustness of our prescription, we compute the Toomre parameter for three additional discs (DM Tau, LkCa 15, and GM Aur) that have published Q profiles \citep[e.g.,][]{Martire_2024, Longarini_2025} and published stellar radii and temperatures \citep{2011Andrews}, enabling us to reproduce their temperature structures within our framework. The resulting Q profiles are presented in Figure \ref{fig:q_r_longarini} and show a similar overall shape, although our profiles display smaller minima of $Q=2-3$ as opposed to their values of $Q=7-8$. This relatively minor discrepancy arises from the different temperature profiles adopted by them (individually derived via modelling), as opposed to those we calculate using Equation \ref{eq:Tprofile}).

\section{Planet Formation Around Post-Common Envelope Binary Systems}
\label{ap:nnser}

For completeness and consistency we calculate here the mass and orbital parameters of the two planets of NN Ser using a realistic temperature profile that allows us not to assume a value of $Q$, and compare our values with those of \citet{2014schleicher}.

\citet{2014schleicher} started  their analysis of the NN~Ser post-common envelope (CE) binary and its putative planets, by motivating their choice of disc mass ($\sim 0.1$~\msun{}).

They then estimated the amount of angular momentum deposited by the companion into the envelope of the AGB star, $L_{\rm dep}$:
\begin{equation}
   \label{eq:angmom_dep}
   L_{\rm dep}=M_2R_*\sqrt{\frac{GM_1}{R_*}}
\end{equation}
where $M_1$ and $R_*$ are the AGB star's mass and radius, respectively and $M_2$ is the companion's mass (see Table~\ref{tab:schleicher_nnser}). In order to account for additional angular momentum that the envelope may already have had at the time of CE in-spiral, they assumed that the actual disc's specific angular momentum would be a factor of a few larger than the one deposited by the companion.  They parameterised this in the following way:
\begin{equation}
   \label{eq:alphal}
   \frac{L_{\rm disc}}{M_{\rm disc}}=\alpha_{\rm L} \frac{L_{\rm dep}}{M_{\rm ej}}
\end{equation}
where $\alpha_{\rm L}$ is the enhancement factor for the specific angular momentum, $L_{\rm disc}$ is the angular momentum of the disc and $M_{\rm ej}$ is the mass ejected during the CE phase, but that presumably remains bound and forms the disc.

After the disc has formed, they assumed that its surface density, $\Sigma$, settles into a power-law radial profile with a truncation radius, $r_{\rm out}$, described by Equation~\ref{eq:surface_density_power_law}.
If they calculate the angular momentum of the disc assuming Keplerian velocity surrounding a stripped core with mass $M_{\rm c}$ and a companion star of mass $M_2$, they could find the outer radius, $r_{\rm out}$:
\begin{equation}
    \label{eq:routschleicher}
    r_{\rm out}=\left(\frac{2.5-n}{2-n}\right)^2\left(\frac{L_{\rm disc}}{M_{\rm disc}}\right)^2\frac{1}{G(M_{\rm c}+M_2)}, 
\end{equation}
with $L_{\rm disc}/M_{\rm disc}$  from Equation \ref{eq:alphal}. 

\begin{table}[hbt!]
\begin{threeparttable}
\caption{Parameters adopted or determined by \citet{2014schleicher} in their study of NN~Ser.}
\label{tab:schleicher_nnser}
\resizebox{\columnwidth}{!}{
\begin{tabular}{ll}
\toprule
\headrow Quantity  & Value  \\
\midrule
Disc Mass, $M_{\rm disk}$ (M$_\odot$) & 0.146\\ 
Ejected Mass, $M_{\rm ej}$ (M~$_\odot$)& 1.24\\
Primary mass, $M_1$ (M~$_\odot$ )& 2 \\
Companion mass, $M_2$ (M~$_\odot$) & 0.111\\ 
Primary core mass, $M_{\rm c}$ (M~$_\odot$ )& 0.535 \\
Primary radius, $R_*$ (R$_\odot$)& 185 \\
Disc's surface density index, $n$ & 1\\ 
Disc's outer radius, $r_{\rm out}$ (R$_\odot$) & 7.5 \footnote{\label{note:diff}The value we find for $r_{\rm out}$ and $L_{\rm dep}$ is different from that of \citet{2014schleicher}, see text for more details.} \\
Disc's ang. mom., $L_{\rm dep}$ (g~cm~s$^{-1}$) & $1.2\times10^{52}$ \footref{note:diff}\\
Disc's ang. mom. enhancement factor, $\alpha_{\rm L}$ & 12.5 \\
critical viscosity, $\alpha_{\rm crit}$ & 0.3 \\
\bottomrule
\end{tabular}}
\end{threeparttable}
\end{table}

\citet{2014schleicher} made the assumption that this disc is gravitationally unstable ($Q=1$). From an initial clump mass, $M_{\rm cl}$, approximately estimated as: 

\begin{equation}
\label{eq:clump}
M_{\rm cl} (r) \approx \Sigma(r)h(r)^2,
\end{equation}
the clumps will then grow on a dynamical timescale to a final mass: 
\begin{equation}
\label{eq:final_clump}
    M_{\rm f}(r)=M_{\rm cl} (r)\left(12\pi^2\frac{\alpha_{\rm crit}}{0.3}\frac{r}{H(r)}\right)^{0.5},
\end{equation}
with $\alpha_{\rm crit}$ the critical viscosity parameter. This equation is obtained from the gap-opening mass expression of \citet{1986lin}\footnote{Equation 34 of \citet{2014schleicher} has a missing factor of $\sqrt{\pi}$ compared to that of \citet{1986lin}.}. They obtained masses of 6.6~M$_{\rm Jup}$ and 1.9~M$_{\rm Jup}$ (cf. observed values of 7~M$_{\rm Jup}$ and 1.7~M$_{\rm Jup}$) for orbital separations of 5.4~au and 3.4~au.

We have summarised the values used by \citet{2014schleicher} in Table~\ref{tab:schleicher_nnser}. Using these values, we found $L_{\rm dep} =1.3\times10^{52}$~erg~cm$^2$~s$^{-1}$ and $r_{\rm out}=7.50$~au  (while  \citet{2014schleicher} reported $1.2\times10^{52}$~erg~cm$^2$~s$^{-1}$ and $r_{\rm out}=6.48$~au).

We then estimate the final clump mass (Equations \ref{eq:clump} and \ref{eq:final_clump}) by using a realistic temperature profile and not assuming that $Q=1$ (Equation \ref{eq:Tprofile}).
The predicted final clump masses for orbital separations of 5.4~au and 3.4~au (blue line) are 2~M$_{\rm Jup}$ and 1.2~M$_{\rm Jup}$, somewhat lower than the values obtained by \citet{2014schleicher}. Despite this small discrepancy, the analysis of \citet{2014schleicher} is consistent with what we have found, so that we can conclude that if the disc mass is close to their adopted 0.1~\msun, then indeed this disc could make the observed planets by gravitational fragmentation.

\begin{figure}[hbt!]
\centering
\includegraphics[width=\linewidth]{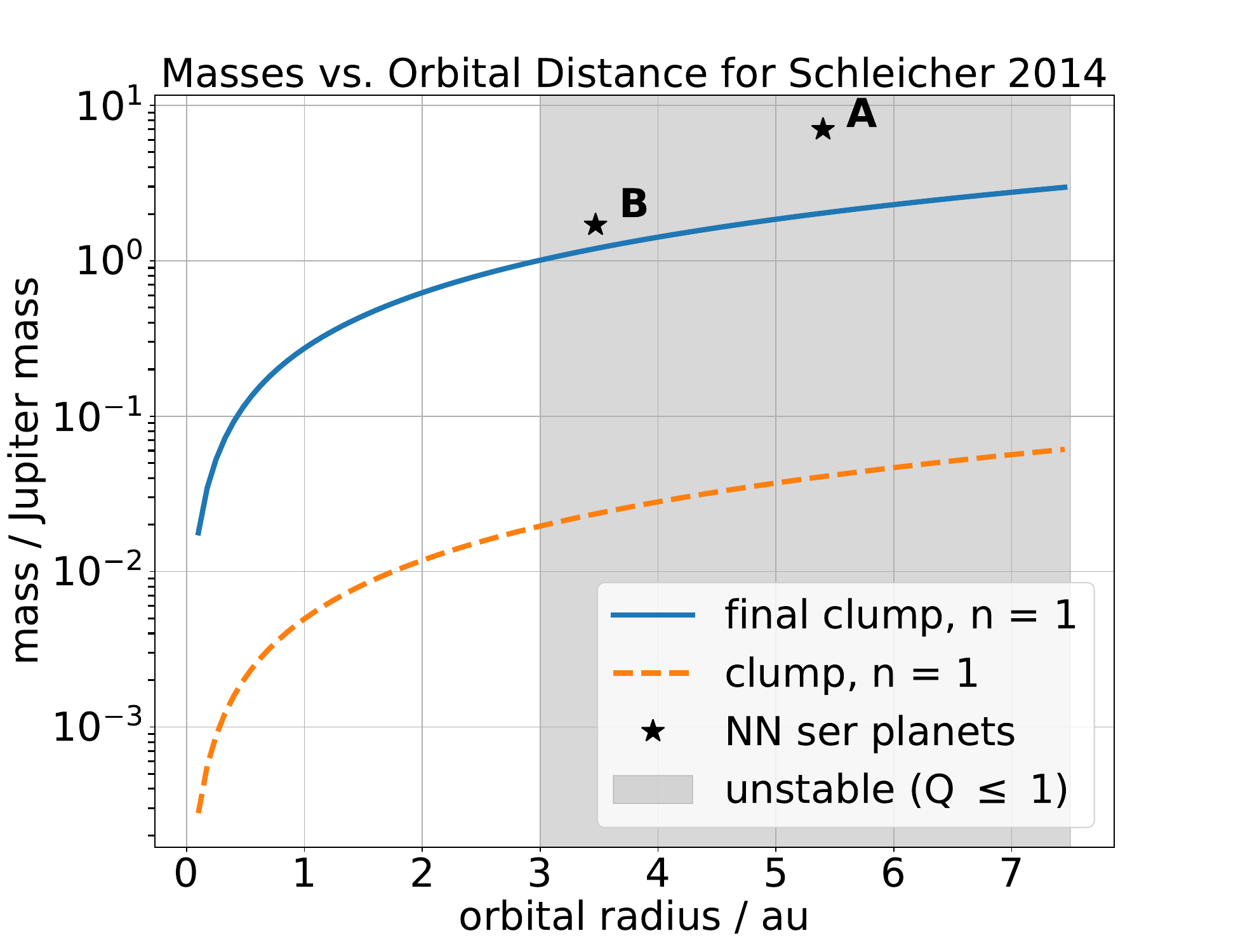}
\caption{Clump masses and final (planet) masses as a function of orbital distance for NN~Ser using the values used by \citet{2014schleicher}, but assuming the  temperature profile of Equation~\ref{eq:Tprofile} and $r_{\rm out}=7.5$~au. }
\label{fig:Mvsr_schleicher}
\end{figure}

\end{document}